\documentclass[12pt,a4paper]{article}

\usepackage{amsmath,amsfonts,latexsym,amssymb}
\usepackage{verbatim,amsthm,curves,graphics}
\usepackage{mathrsfs}

\theoremstyle{definition}
\newtheorem{thm}{Theorem}[section]

\newtheorem{lemma}{Lemma}[section]
\newtheorem{prop}{Proposition}[section]

\begin{document}

\renewcommand{\j}{{r}}
\newcommand{\mye}{\stackrel{\circ}{e}}
\newcommand{\myn}{\stackrel{\circ}{\partial}}
\newcommand{\myI}{\stackrel{\circ}{I}}
\newcommand{\mr}{\mathbb{R}} 
\newcommand{\mz}{\mathbb{Z}} 
\newcommand{\mc}{\mathbb{C}} 
\newcommand{\mh}{\mathbb{H}} 
\newcommand{\ms}{\mathbb{S}} 
\newcommand{\mn}{\mathbb{N}} 
\newcommand{\F}{{\bf F}} 
\newcommand{\beps}{\mbox{\boldmath $\epsilon$}} 
\newcommand{\pgator}{/\!\!\! S}
\newcommand{\rP}{\operatorname{P}}
\renewcommand{\i}{{\rm i}}
\newcommand{\car}{\operatorname{CAR}}
\newcommand{\rSpin}{\operatorname{Spin}}  
\newcommand{\rSO}{\operatorname{SO}}      
\newcommand{\rO}{\operatorname{O}}        
\newcommand{\rCliff}{\operatorname{Cliff}}
\newcommand{\WF}{\operatorname{WF}}       
\newcommand{\Pol}{{\rm WF}_{pol}}         
\newcommand{\cC}{\mathscr{C}}             
\newcommand{\cA}{\mathcal{A}}                
\newcommand{\cW}{\mathcal{W}}                
\newcommand{\cX}{\mathcal{X}}                
\newcommand{\cO}{\psi}
\newcommand{\bO}{\mbox{\boldmath $O$}}
\newcommand{\Star}{\operatorname{\mbox{\huge $\star$}}}
\newcommand{\cV}{\mathcal{V}}             
\newcommand{\ccR}{\mathcal{R}}             
\newcommand{\cS}{\mathcal{S}}             
\newcommand{\cF}{\mathcal{F}}             
\newcommand{\cE}{\mathcal{E}}             
\newcommand{\cB}{\mathcal{B}}
\newcommand{\cK}{\mbox{\boldmath $K$}}
\newcommand{\cP}{\mathcal{P}}
\newcommand{\cD}{\mathcal{D}_1}             
\newcommand{\cL}{{\mbox{\boldmath $L$}}}             
\newcommand{\cQ}{\mathscr{Q}}             
\newcommand{\ccS}{\mathscr{S}}             
\newcommand{\bS}{{\mbox{\boldmath $\mathcal{S}$}}}
\newcommand{\tbS}{{\mbox{\boldmath $\widetilde{\mathcal{S}}$}}}
\newcommand{\bF}{\mbox{\boldmath $F$}}
\newcommand{\bdel}{\mbox{\boldmath $\delta$}}
\newcommand{\bU}{\mathcal{U}}
\newcommand{\bL}{\operatorname{OP}}
\newcommand{\glob}{{\textrm{\small global}}}
\newcommand{\loca}{{\textrm{\small local}}}
\newcommand{\singsupp}{\operatorname{singsupp}}
\newcommand{\dom}{\operatorname{dom}}
\newcommand{\clo}{\operatorname{clo}}
\newcommand{\supp}{\operatorname{supp}}
\newcommand{\rd}{{\rm d}}                 
\newcommand{\mslash}{/\!\!\!}             
\newcommand{\slom}{/\!\!\!\omega}         
\newcommand{\dirac}{/\!\!\!\nabla}        
\newcommand{\myid}{\leavevmode\hbox{\rm\small1\kern-3.8pt\normalsize1}}
\newcommand{\esssup}{\operatorname*{ess.sup}}
\newcommand{\ran}{\operatorname{ran}}
\newcommand{\sd}{{\rm sd}}
\newcommand{\reg}{\,{\rm l.n.o.}}
\newcommand{\isom}{\iota}
\newcommand{\wk}{{\bf k}}
\newcommand{\ws}{{\bf s}}
\newcommand{\lno}{:\!}
\newcommand{\rno}{\!:}
\newcommand{\clim}{\operatorname*{coin.lim}}
\newcommand{\mydef}{\stackrel{\textrm{def}}{=}}
\renewcommand{\min}{{\textrm{\small int}\,\mathscr{L}}}
\newcommand{\sym}{\operatorname{sym}}
\renewcommand{\Im}{\operatorname{Im}}
\renewcommand{\Re}{\operatorname{Re}}
\newcommand{\Exp}{\operatorname{Exp}}
\newcommand{\Ad}{\operatorname{Ad}}
\newcommand{\sa}{\mathfrak{sa} } 
\newcommand{\so}{\mathfrak{so} } 
\renewcommand{\o}{\mathfrak{o} } 
\newcommand{\su}{\mathfrak{su} } 
\newcommand{\sq}{\mathfrak{sq} } 
\renewcommand{\sp}{\mathfrak{sp} } 
\newcommand{\gl}{\mathfrak{gl}} 
\newcommand{\g}{{\bf g}}	
\newcommand{\bLa}{\mbox{\boldmath $\Lambda$}}	
\newcommand{\h}{\mathfrak{h} } 
\newcommand{\f}{\mathfrak{f} } 
\newcommand{\p}{\mathfrak{p} } 
\newcommand{\U}{\operatorname{U}} 
\newcommand{\Z}{\operatorname{Z}} 
\newcommand{\SO}{\operatorname{SO} } 
\newcommand{\SU}{\operatorname{SU} } 
\renewcommand{\O}{\operatorname{O} }
\newcommand{\SP}{\operatorname{Sp} } 
\newcommand{\SL}{\operatorname{SL}} 
\newcommand{\GL}{\operatorname{GL}} 
\newcommand{\Der}{\operatorname{Der}} 
\newcommand{\Str}{\operatorname{Str}} 
\newcommand{\End}{\operatorname{End}} 
\newcommand{\Cl}{\operatorname{Cl}} 
\newcommand{\ds}{\dot{+}}
\renewcommand{\aa}{\alpha}
\renewcommand{\ln}{\log}
\newcommand{\lnabla}{{\stackrel{\leftarrow}{\nabla}}}
\newcommand{\rnabla}{{\stackrel{\rightarrow}{\nabla}}}
\newcommand{\x}{\bar{x} } 
\newcommand{\y}{\bar{y} }
\newcommand{\s}{\bar{s} }
\newcommand{\diag}{\operatorname{diag}}
\renewcommand{\Ad}{\operatorname{Ad}}
\newcommand{\sgn}{\operatorname{sgn}}
\newcommand{\tr}{\operatorname{tr}} 
\newcommand{\per}{\operatorname{per}} 
\newcommand{\Tri}{\operatorname{Tri}}
\renewcommand{\square}{\nabla^a \nabla_a}

\title{On the Renormalization Group in Curved Spacetime}
\author{Stefan Hollands\thanks{Electronic mail: \tt stefan@bert.uchicago.edu}
        \,\, and
        Robert M. Wald\thanks{Electronic mail: \tt rmwa@midway.uchicago.edu}\\
       \it{Enrico Fermi Institute, Department of Physics,}\\
       \it{University of Chicago, 5640 Ellis Ave.,}\\
       \it{Chicago IL 60637, USA}
        }
\date{\today}

\maketitle

\begin{abstract}
We define the renormalization group flow for a renormalizable
interacting quantum field in curved spacetime via its behavior under
scaling of the spacetime metric, $\g \rightarrow \lambda^2 \g$. We
consider explicitly the case of a scalar field, $\varphi$, with a
self-interaction of the form $\kappa \varphi^4$, although our results
should generalize straightforwardly to other renormalizable
theories. We construct the interacting field---as well as its Wick
powers and their time-ordered-products---as formal power series in the
algebra generated by the Wick powers and time-ordered-products of the
free field, and we determine the changes in the interacting field
observables resulting from changes in the renormalization
prescription. Our main result is the proof that, for any fixed
renormalization prescription, the interacting field algebra for the
spacetime $(M, \lambda^2 \g)$ with coupling parameters $p$ is
isomorphic to the interacting field algebra for the spacetime $(M,
\g)$ but with different values, $p(\lambda)$, of the coupling
parameters. The map $p \to p(\lambda)$ yields the renormalization
group flow. The notion of essential and inessential coupling
parameters is defined, and we define the notion of a fixed point as a
point, $p$, in the parameter space for which there is no change in essential
parameters under renormalization group flow.

\end{abstract}

\pagebreak

\section{Introduction}

Theories of a classical field in Minkowski spacetime that are derived
from an action principle will automatically possess an invariance
under a scaling of the global inertial coordinates of spacetime (or,
equivalently, under scaling of the field momenta) provided that a
corresponding scaling of the field amplitude and coupling constants
are also performed in such a way that the action remains unchanged. If the
quantum theory of this field is renormalizable, it turns out that in
perturbation theory there also is a similar invariance of quantities
of interest---such as the Green's functions of the fields---under
scaling of the field momenta, but the required scaling of the field
amplitudes and coupling constants differs, in general, from the simple
scaling laws for the classical theory. This change of the ``field
strength normalization'' and coupling constants under scaling is
called the ``renormalization group flow'' of the theory. Important
qualitative as well as quantitative information about quantum field
theories can be gained from an analysis of their renormalization group
flow.

For quantum field theories in Minkowski spacetime, there exist well
known procedures for calculating the renormalization group flow in
perturbation theory.  In many cases, the picture obtained from low
orders is believed to be at least in qualitative agreement with the
behavior that would hold in the full, nonperturbatively constructed
quantum field theory. Consequently, perturbative calculations of the
renormalization group flow have played an important role in arguments
concerning fundamental properties of quantum field theories. In
particular, they form the basis of the claim that certain non-abelian
gauge theories are ``asymptotically free'', i.e., that the gauge
coupling flows towards zero at small distances (large momenta).

It is therefore of interest to know whether a similar scaling analysis
can also be performed for perturbative interacting quantum field
theory on an arbitrary globally hyperbolic curved (Lorentzian)
spacetime. As we shall briefly review in section 2 below, the
construction of perturbative interacting quantum field theory in
curved spacetime has recently been achieved in \cite{hw1},
\cite{hw2}, based upon some earlier key results established in
\cite{bfk, bf1} and other references. However, for at least the
following two reasons, it does not seem possible to give a
straightforward generalization to curved spacetime of the usual
scaling analyses given for Minkowski spacetime. First, as already
indicated above, the renormalization group flow in Minkowski spacetime
is usually formulated in terms of behavior under the scaling of
global inertial coordinates or, equivalently, scaling of the field
momenta. However, in curved spacetime a formulation in terms of
scaling of coordinates (or momenta) would introduce a very awkward and
undesired coordinate dependence into the constructions. Also, since
the scaling of coordinates no longer corresponds to a conformal
isometry of the spacetime metric, one would not expect a simple
behavior to occur under scalings of any coordinates.  Second, the
quantities whose scaling behavior is usually considered in studying
the renormalization group flow in Minkowski space are the Green's
functions of the interacting field or other quantities from which
these can be derived, such as the ``effective action''. However, the
Green's functions depend on a choice of state. For quantum field
theories in Minkowski spacetime, this state would naturally be chosen
to be the (unique) Poincare invariant vacuum state. However, even for
a free quantum field in a general curved spacetime, there is no
``preferred vacuum state'' nor any other state that can be singled out
for special consideration. Thus, even if a renormalization group flow
could be defined in terms of Green's functions, there is no reason to
expect it to be independent of the choice of state used to define the
Green's functions.

A solution to the second difficulty is achieved by formulating the
theory via the algebraic approach. In this approach, one views the
observables as forming an abstract algebra, and one views the quantum
states as suitable linear functionals on this algebra. This algebra is
referred to as ``abstract'', because no representation of this algebra
on a particular Hilbert space has been chosen from the outset, so that
the (potentially problematic) issue of choosing states is completely
disentagled from the issue of constructing the observables of the
theory. As we shall see, the renormalization group flow can then be
defined at the level of the algebra of observables.

The first difficulty above is solved by defining the renormalization
group flow in terms of the behavior of the algebra of the interacting
field under a scaling of the spacetime metric, $\g \to \lambda^2 \g$,
as has previously been suggested by other authors~\cite{pt,tj1,tj2}. 
In Minkowski spacetime, the diffeomorphism
defined by the rescaling of the global inertial coordinates, $x^\mu
\to \lambda x^\mu$, is a conformal isometry with constant conformal
factor $\lambda$, so rescaling the coordinates or momenta is equivalent to
rescaling the spacetime metric. However, in a general curved spacetime
there will not exist any conformal isometries, so rescaling the metric
is not equivalent to any rescaling of coordinates or momenta. As we
shall see, in perturbation theory the interacting field has a well
defined behavior under scaling of the spacetime metric.

The results we shall obtain in this paper are based primarily on our
previous uniqueness theorems \cite{hw1} for Wick polynomials and their
time-ordered products for a free quantum field. As we shall explain
further in section 2 below, these results imply that the interacting
field algebra is well defined up to certain renormalization
ambiguities. In particular, for the case of a renormalizable theory,
the ambiguities in the interacting field algebra correspond precisely
to changes in the (finite number of) parameters appearing in the
interaction Lagrangian\footnote{In other words, if one changes the
prescription for defining Wick products and their time ordered
products for the free theory in a manner compatible with the axioms of
\cite{hw1} and \cite{hw2}, the new interacting field algebra one
obtains via the construction given in section 3 below will correspond
to the interacting field algebra obtained with the original
prescription, but with the interaction Lagrangian modified by the
addition of terms of the same form as appearing in the original
Lagrangian. The definition of the interacting field with the new
prescription will also correspond up to a numerical factor to the
definition of the interacting field in the corresponding algebra
obtained from the original prescription with the modified Lagrangian,
i.e., the isomorphism of the interacting field algebras for the two
different prescriptions will map the interacting field to a multiple
of the interacting field. It should be noted, however, that the new
definition of higher Wick powers of the interacting field (as well as
time-ordered-products of Wick powers of the interacting field) will
not correspond to the definition of these quantities obtained from the
original prescription with the modified Lagrangian. Instead, under the
isomorphism of the algebras, a higher Wick power (or a
time-ordered-product of Wick powers) will, in general, be mapped into
a field of the form specified in eq.~(\ref{VAV}) below.}. This
observation gives rise to the following means to define the
renormalization group flow: Fix a renormalization prescription for
defining the free field Wick polynomials and their time ordered
products. Now apply this renormalization prescription to define Wick
polynomials and their time-ordered-products for free quantum fields on
the spacetime $(M, \lambda^2 \g)$, with all of the parameters of the
theory also scaled according to their ``engineering dimension'' (i.e.,
scaled in such a way as to keep the classical action invariant). The
free field algebra of observables $\cW(M, \g)$ (defined in \cite{hw1}
and in section 2 below) is naturally isomorphic to $\cW(M, \lambda^2
\g)$ with scaled parameters, and we can use this isomorphism to define
a new ($\lambda$-dependent) renormalization prescription for Wick
polynomials and their time-ordered products on the original spacetime
$(M, \g)$. We thereby obtain a new ($\lambda$-dependent) prescription
for defining the interacting field algebra. However, by our uniqueness
results, this prescription must be equivalent to the original
prescription for defining the interacting field algebra modulo a
change of parameters appearing in the interaction
Lagrangian. Consequently, we get a $\lambda$-dependent ``flow'' in the
parameter space of the interacting theory\footnote{In other words, if
we scale the spacetime metric and correspondingly scale the
parameters, $p_0$, of the free Lagrangian, $\cL_0$, according to their
``engineering dimension'', then the resulting theory is equivalent to
a theory where the metric and parameters, $p_0$, are not scaled, but
the {\em interaction} Lagrangian, $\cL_1$, is modified by
$\lambda$-dependent terms of the same form as appear in the (full)
Lagrangian $\cL = \cL_0 + \cL_1$. It should be emphasized that it
is far from obvious that, for a perturbatively constructed
interacting theory, a change in a parameter appearing in $\cL_1$ as
occurs in the renormalization group flow is equivalent to
a corresponding change in that parameter in $\cL_0$; see the end of
section 4.1 for further discussion.}. This flow defines the action of
the renormalization group for a quantum field in curved spacetime.

In order to implement the above ideas, we first must define the
interacting quantum field algebra and therefore must address the
following two difficulties: (i) As in Minkowski spacetime, the
interacting quantum field is defined only perturbatively, and it is
not expected that the perturbation series converges. (ii) The usual
formula for defining the interacting field expresses it in terms of a
free ``in''-field~\cite{haag}. Even if the theory under consideration is
such that in Minkowski spacetime the interacting field approaches a 
free ``in''-field in the asymptotic past in a suitable sense, 
there is no reason to expect any such behavior to
occur in an arbitrary globally hyperbolic curved spacetime.

As discussed in section 3.1, we shall, in essence, sidestep issue (i) by
treating the interacting field algebra only at the level of a formal
perturbation series. In other words, we do not attempt to define the
interacting field algebra at a finite value of a nonlinear coupling
parameter, $\kappa$, but simply consider the algebra generated by the
formal perturbation series expressions in $\kappa$. In this respect,
our analysis is neither better nor worse than the corresponding
analyses for perturbative quantum field theory in Minkowski
spacetime. We note, however, that at least some of the difficulties
encountered in making sense of perturbative expansions for nonlinear
quantum field theory may be due to the non-analytic behavior of ground
states and/or ``in'' and ``out'' states. It appears conceivable that
at least some of the difficulties of perturbation theory could be averted if one works
strictly at the algebraic level and uses perturbation formulas only to
obtain algebraic relationships between interacting field observables
(thereby defining the interacting field algebra) rather than using
perturbation theory to calculate quantities involving, say, ground
states or ``in'' and ``out'' states. However, we shall not attempt to
pursue these ideas in this paper.

On the other hand, difficulty (ii) can be genuinely overcome by
properly taking limits as the cutoff on the interaction is removed:
The Bogoliubov formula defining the interacting field (see
eq.~(\ref{intfield}) below) is well defined if the nonlinear coupling
parameter, $\kappa$, is taken to be a smooth function of compact
support, so that the nonlinear interaction is ``turned off'' in the
past and future. If one then attempts to take a limit where $\kappa$
approaches a constant, difficulties may arise if one demands that the
interacting field remain fixed in, say, the asymptotic past. However,
no difficulties arise if, following the ideas of \cite{bf1}, we demand
that the interacting field remain fixed in the ``interior'' of the
spacetime as $\kappa$ approaches a constant. This construction is
given in section 3.1.

The organization of this paper is as follows. In section 2, we briefly
review the main ingredients that we will need from free quantum field
theory in curved spacetime, including the definition and uniqueness
properties of Wick powers and their time-ordered-products. In section
3.1 we give the construction of the interacting field and in section
3.2 we characterize its renormalization ambiguities. The scaling
behavior of the interacting theory is analyzed in section 4.1, and the
renormalization group flow is defined. The notions of essential and
inessential coupling parameters and the notion of ``fixed points''
under the renormalization group flow are defined in section 4.2. In
appendix B, we will relate our rather abstract formulation of
renormalization theory and the renormalization group flow 
at the algebraic level to more usual formulations in terms of Feynman diagrams.

In this paper, we will consider only a scalar field with Lagrangian
density of the form
\begin{equation}
\label{L}
\cL = \cL_0 + \cL_1 \equiv 
\frac{1}{2}[(\nabla\varphi)^2 + m^2 \varphi^2 + 
\xi R \varphi^2 + \kappa\varphi^4]\beps, 
\end{equation} 
where, $R$ is the scalar curvature and $\beps$ is the volume element
constructed from the spacetime metric $\g = \g_{ab}$. 
The self-interaction $\cL_1 = \frac{1}{2}\kappa\varphi^4 \beps$
will be treated perturbatively. However, all of our analysis should
generalize straightforwardly to other renormalizable quantum field
theories. 

Our notation and conventions follow those of our previous papers
\cite{hw1}, \cite{hw2}. All spacetimes $(M,\g)$ considered in this
paper will be assumed to be globally hyperbolic and time oriented. We
will denote the free quantum scalar field (defined by the Lagrangian
(\ref{L}) with $\kappa = 0$) by $\varphi$ and will use the generic
notation $\Phi$ to denote other local covariant fields in the free
theory. The interacting field will be denoted $\varphi_{L_1}$ and
other local covariant fields in the interacting theory will be denoted
$\Phi_{L_1}$. In this paper, all fields will be smeared with scalar 
{\em densities} (of unit weight); we will denote the space of smooth 
unit weight scalar densities of compact support on $M$ by $\cD(M)$.

\section{The free quantum field in curved spacetime}

The perturbative construction of a self-interacting quantum scalar
field in curved spacetime is based upon the construction of the free
quantum field theory. In this section, we consider the quantum field
theory of a free scalar field $\varphi$, described by the classical
Lagrangian density
\begin{equation}
\label{Lfree}
\cL_0 = \frac{1}{2}[(\nabla \varphi)^2 + m^2 \varphi^2 + 
\xi R \varphi^2]\beps.  
\end{equation}
Note that under a scaling
of metric, $\g \to \lambda^2 \g$ with $\lambda$ a positive constant,
the Lagrangian density remains invariant provided that we also scale
the field, $\varphi$, mass, $m$, and coupling parameter $\xi$, by
$\varphi \to \lambda^{-1} \varphi$, $m \to \lambda^{-1} m$, $\xi \to
\xi$. We refer to the power of $\lambda$ appearing in these scaling
rules as the {\it engineering dimension} of the quantity. More
generally, any monomial, $\Phi$, constructed out of $\varphi$ and its
derivatives, the curvature, and the coupling constants $m$ and $\xi$
will have a well defined engineering dimension, denoted $d_\Phi$.

As is well known, in a general curved spacetime, there is no
``preferred vacuum state'' nor even any preferred Hilbert space
construction of the quantum theory corresponding to the classical
Lagrangian (\ref{Lfree}) (see, e.g., \cite{wa} for further
discussion).  Therefore, in our view, it is essential to formulate the
theory via the algebraic approach.

As in \cite{hw1}, we shall take the algebra of observables of the free
field to be the ``extended Wick polynomial algebra''
$\cW(M,\g)$. As described in \cite{hw1}, this algebra can be
constructed by choosing a quasifree Hadamard state, $\omega$, on the
``canonical commutation algebra'', $\cA(M,\g)$, then considering the
normal ordered field operators on the GNS representation of $\omega$,
and showing \cite{bfk} that one gets well defined operators by
smearing these normal ordered operators with suitable distributions
rather than test functions. The resulting algebra of operators can
then be shown \cite{hw1} to be independent of the choice of $\omega$.

Following \cite{fd}, we outline here a much more direct construction
of $\cW(M,\g)$. This construction is sufficiently different in
appearance from that given in \cite{hw1} that it is worthwhile to
explain the relationship between the constructions. First, recall the
usual construction of the canonical commutation algebra, $\cA(M,\g)$:
Start with the free *-algebra generated by the identity, $\myid$, and all
expressions of the form $\varphi(f)$, where $f$ is an element of 
$\cD(M)$, the space of smooth scalar densities on $M$ with compact support.
(Thus, this algebra consists of all finite linear combinations of $\myid$ and
terms containing finitely many factors of the form $\varphi(f_i)$ and
$\varphi(f_j)^*$.) Next, define the two-sided ideal consisting of all
elements of this algebra that contain at least one factor of any of
the following four types: 
\begin{enumerate}
\item[(i)]
$\varphi(\alpha_1 f_1 + \alpha_2 f_2) - \alpha_1 \varphi(f_1) - \alpha_2
\varphi(f_2)$, with $\alpha_1, \alpha_2
\in \mc$; 
\item[(ii)]
$\varphi(f)^* - \varphi(\bar{f})$; 
\item[(iii)]
$\varphi((\square - m^2  - \xi R)f)$; and 
\item[(iv)]
$\varphi(f_1) \varphi(f_2) -\varphi(f_2) \varphi(f_1) - i\Delta(f_1,
f_2) \myid$, where $\Delta$ denotes the advanced minus retarded Green's
function for the Klein-Gordon operator. 
\end{enumerate}
Then $\cA(M,\g)$ is defined by factoring the free algebra by
this ideal.

It is useful to make the following trivial change in the
construction of $\cA(M,\g)$: Instead of starting with the free algebra
generated by the identity, $\myid$, and symbols of the form $\varphi(f)$,
we start with the free tensor algebra of smooth compactly supported scalar test densities on $M$,
\begin{equation}
\label{test}
\cF(M) \equiv \mc \oplus \bigoplus_{n \ge 1} \otimes^n \cD (M). 
\end{equation}
with a *-operation defined by complex conjugation. (Note that although
the direct sum in eq.~(\ref{test}) is infinite, by definition, each
element of $\cF(M)$ has only finitely many non-zero entries.) The
*-algebra $\cF(M)$ already incorporates the identifications
corresponding to (i) and (ii) above, and clearly is isomorphic to the
free algebra of the previous paragraph factored by the ideal generated
by (i) and (ii). Thus, we can equivalently define $\cA(M,\g)$ by
factoring $\cF(M)$ by the ideal generated by expressions (iii) and
(iv) above. We will incorporate this viewpoint in our notation by
denoting elements of $\cA(M,\g)$ by their representatives in
$\cF(M)$. Thus, for example, we will denote the element of $\cA(M,\g)$
corresponding to the field operator smeared with $f \in \cD(M)$ by $[f]$
rather than $\varphi(f)$.

Next, we note that given any $t \in \cF(M)$, the imposition of the
commutation relations (iv) above would allow us to choose a unique
representative of $t$ in the totally symmetric tensor algebra. Thus,
rather than imposing these commutation relations by factorization as
above, we may instead work with the totally symmetric tensor
algebra. Hence, we define
\begin{equation}
\label{-}
\cF_{\sym}(M) \equiv \mc \oplus \bigoplus_{n \ge 1}
\otimes_{\sym}^n \cD(M).
\end{equation}
and we define a product, $\star_0$, (which depends upon $\g$) in
$\cF_{\sym}(M)$ that corresponds to taking the ordinary tensor
product in $\cF(M)$. Namely, if $t_n \in \otimes^n_{\sym} \cD(M)$
and $s_m \in \otimes_{\sym}^m \cD(M)$, we define
\begin{multline}
(t_n \star_0 s_m)_{n+m-2k}(x_1, \dots, x_{n+m-2k}) = \frac{n!m!}{k!
(n-k)!(m-k)!}
{\mathbb S} \int_{M^{2k}} 
t_n(y_1, \dots, y_k, x_1, \dots, x_{n-k}) \\
s_m(y_{k+1}, \dots, 
y_{2k}, x_{n-k+1}, \dots, x_{n+m-2k})
\prod_{i=1}^k \frac{i}{2}\Delta(y_i, y_{k+i}),
\label{symprod0}  
\end{multline}
where ``${\mathbb S}$'' denotes total symmetrization in the variables
$x_1, \dots, x_{n+m-2k}$ and where the integral is over the ``$y$''-variables\footnote{
Since $t_n$ and $s_m$ are densities, no volume element has to be specified in 
the integral.}. In other words, the right side of eq.~(\ref{symprod0}) gives the
totally symmetric representative of $t_n \otimes s_m$ in the tensor
algebra $\cF(M)$ under imposition of the commutation relations
(iv). Since the algebra (\ref{-}) with the product (\ref{symprod0})
already incorporates conditions (i), (ii), and (iv) above, we consider
the ideal consisting of all elements of $\cF_{\sym}(M)$
that contain at least one factor of the form $(\square -
m^2  - \xi R )f$. We again obtain $\cA(M,\g)$ by factoring $\cF_{\sym}(M)$ by 
this ideal.

We now make an important further modification to the above
construction by introducing a new ($\omega$-dependent) product, $\star$, on
$\cF_{\sym}(M)$ by replacing $\frac{i}{2}\Delta$ in eq.~(\ref{symprod0}) by
$\omega$ where $\omega$ is an arbitrary (``undensitized'') distribution in two variables
that satisfies the Klein-Gordon equation in each variable and whose
antisymmetric part is equal to $\frac{i}{2}\Delta$, 
\begin{multline}
(t_n \star s_m)_{n+m-2k}(x_1, \dots, x_{n+m-2k}) = \frac{n!m!}{k!
(n-k)!(m-k)!}
{\mathbb S} \int_{M^{2k}} 
t_n(y_1, \dots, y_k, x_1, \dots, x_{n-k}) \\
s_m(y_{k+1}, \dots, 
y_{2k}, x_{n-k+1}, \dots, x_{n+m-2k})
\prod_{i=1}^k \omega(y_i, y_{k+i}), 
\label{omegaprod}  
\end{multline}
where the integral is again over the ``$y$''-variables.
Then, by the same argument as in Lemma~2.1 of \cite{hw1}, 
it can be seen that $\cF_{\sym}(M)$ with the
product $\star$ is naturally isomorphic to $\cF_{\sym}(M)$ with the
product $\star_0$. Therefore if we factor $\cF_{\sym}(M)$ with the product
$\star$ by the ideal comprised by all elements of $\cF_{\sym}(M)$
that contain at least one factor of $(\square - m^2  -
\xi R)f$, we again obtain an algebra isomorphic to $\cA(M,\g)$.
It also should be noted that for $f_1, f_2 \in \cD(M)$ we have
\begin{equation}
f_1 \star f_2 - f_2 \star f_1 = i\Delta(f_1, f_2) 
\myid. 
\end{equation}

Now, choose $\omega$ to be the two-point function of a Hadamard
state. Then the product (\ref{omegaprod}) corresponds to
Wick's formula expressing the product of a normal-ordered $n$-point
function with a normal ordered $m$-point function in terms of normal
ordered products, where the normal ordering is done with respect to
the quasi-free Hadamard state with two-point function $\omega$. 
It can thereby be seen that for any $t_n \in \otimes^n_{\sym} \cD(M)$
of the form $t_n = f_1 \otimes_{\sym} \cdots \otimes_{\sym} f_n$
with each $f_i \in \cD(M)$, the algebraic element $[t_n] \in \cA(M, \g)$ 
corresponding to the equivalence class of $t_n$ is represented by 
the normal ordered product $\lno \varphi(f_1) \cdots \varphi(f_n) \rno_\omega$
in the GNS-representation of the state $\omega$.

The
key observation needed to define the algebra $\cW(M,\g)$ is to note
that the wavefront set properties of $\omega$ then imply that
eq.~(\ref{omegaprod}) continues to make sense when the test function
space $\otimes^n_{\sym} \cD(M)$ in \eqref{-} is replaced by the much larger
space\footnote{Since the elements in $\cE'_{\sym}(M^{\times n})$ are
distributions, they automatically have the character of densities.
The space $\otimes^n_{\sym} \cD(M)$ can therefore be naturally 
identified with a subspace of $\cE'_{\sym}(M^{\times n})$, without
the need to specify a volume element on $M$.}
\begin{equation}
\cE'_{\sym}(M^{\times n}) = \{ \text{compactly supp. symm. distr. $t_n$} \mid
\WF(t_n) \subset T^*M^n \setminus (V_+^{\times n} \cup V_-^{\times n})\}, 
\label{esymdef}
\end{equation}
where $V_\pm$ is the future/past lightcone with respect to the metric
$\g$, and where ``$\WF$'' denotes the wave-front set of a distribution
\cite{h}. We define $\cW(M,\g)$ to be the vector space
\begin{equation}
\label{--}
\cE'(M, \g) \equiv \mc \oplus \bigoplus_{n \ge 1} \cE'_{\sym}(M^{\times n}). 
\end{equation}
with product (\ref{omegaprod}), factored by the ideal comprised by all
elements of the form $(\square - m^2 - \xi R)_{x_i} t_n(x_1, \dots, x_n)$.
Thus, every element $a \in \cW$ corresponds to an equivalence class
$a = [s]$ of an element $s = s_0 + \sum^n_{k = 1} s_k$, where $s_0 \in \mc$, 
and where $s_k \in \cE'_{\sym}(M^{\times k})$. The product of two elements
in $\cW$ is given by $[s] \star [t] \equiv [s \star t]$. If $f$ is a smooth 
scalar density on $M$ of compact support, then the equivalence class $[f] \in \cW$
corresponds exactly to the smeared free field $\varphi(f)$. 

The definition of the algebra $\cW$ a priori depends on some choice
for $\omega$, but it was shown in \cite{hw1} that different choices
for $\omega$ give rise to isomorphic algebras. Therefore, as an
abstract algebra, $\cW$ is independent of this choice. Since $\cA$ is
naturally a subalgebra of $\cW$, we automatically know what elements
of $\cW$ correspond to the smeared field $\varphi(f)$ and its smeared
$n$-point functions. However, it is not obvious what (if any) elements
of $\cW$ correspond to smeared Wick powers of the field and
time-ordered products of Wick powers.

This issue was addressed in \cite{hw1} and \cite{hw2}, where an
axiomatic approach was taken. A key condition imposed in \cite{hw1}
and \cite{hw2} on the definition of Wick powers and their
time-ordered-products was that they be local, covariant
fields~\cite{bfv}.  In order to define this notion, it is necessary to
think of the fields as being defined not only for a given, fixed
spacetime, but rather for all (globally hyperbolic) spacetimes, and we 
incorporate this viewpoint here by indexing the field with the spacetime
under consideration, such as $\Phi[M, \g]$. If
$(M, \g)$ and $(\tilde M, \tilde \g)$ are two spacetimes such that
there is a causality preserving isometric embedding, $\chi$, of
$(\tilde M, \tilde \g)$ into $(M, \g)$, then the algebra $\cW(\tilde
M, \tilde \g)$ can be regarded as a subalgebra of $\cW(M, \g)$ via a
homomorphism $\alpha_\chi$ in a natural way \cite{hw1}, so that the
free field theory with algebra $\cW(M, \g)$ is a local, covariant
field theory~\cite{bfv}.  The requirement that $\Phi$ be a local covariant field
is then that
\begin{equation}
\label{lcf}
\alpha_\chi(\Phi[\tilde M,\tilde \g](x)) = \Phi[M, \g](\chi(x)).  
\end{equation}

It was shown in \cite{hw1} that this requirement together with a
number of additional requirements (such as commutation properties,
continuity and analyticity conditions, microlocal spectral conditions,
and causal factorization) uniquely determines the definition of Wick
powers and their time-ordered-product up to certain well defined
renormalization ambiguities. Existence of Wick powers satisfying these
properties also was established in \cite{hw1}, and existence of their
time-ordered-products was proven in \cite{hw2}. 

The results of the present paper will rely heavily on the uniqueness
theorem 5.2 of \cite{hw1} for time-ordered-products.  The allowed
ambiguity in the definition of time-ordered-products as given in
theorem 5.2 of \cite{hw1} is rather awkward to state, so we find it
useful to reformulate this theorem in the following manner (see
\cite{boas,fd}). First, we introduce an abstract vector space, $\cV$,
comprised by finite linear combinations of basis elements labeled by
formal products of $\varphi$ and its covariant derivatives,
\begin{equation} 
\cV = {\rm span}_\mc \left\{ \Phi = \prod \nabla_{(a_1} 
\cdots \nabla_{a_i)} \varphi \right\}.
\end{equation}
We refer to the elements of $\cV$ as ``formal'' because we do not 
assume any relations between the fields at this stage. In particular, 
we regard the field and its derivatives as independent quantities which are 
not related by the field equation. Let
\begin{equation}
\cD(M, \cV) \equiv \{\text{smooth densities on $M$ of compact support with
values in $\cV$}\}
\end{equation}
so that an element $F \in \cD(M,\cV)$ can be uniquely expressed as a 
finite sum $F = \sum f_i \Phi_i$ with each $\Phi_i$ a basis element of $\cV$ and
$f_i \in \cD(M)$.  It is convenient to think of a prescription for
defining Wick powers as a linear map from $\cD(M, \cV)$ into the
algebra $\cW(M, \g)$. Thus, a prescription for Wick powers associates
to an element $f(x) \Phi \in \cD(M, \cV)$ an element $\Phi(f) \in
\cW(M, \g)$. Similarly, it is useful to view the $n$-fold time ordered
product of Wick powers as an $n$-times multilinear map
\begin{eqnarray}
T: \mbox{\huge $\times$}^n \cD(M, \cV) &\to& \cW(M, \g)\\
(f_1 \Phi_1,\dots,f_n \Phi_n) &\to& T(\prod \Phi_i(f_i)). 
\end{eqnarray}
The map defining Wick powers is, of course, the special case $n = 1$
of the map defining time-ordered-products.

Let us now suppose that we have two prescriptions for defining
time-ordered-products (and, in particular, two prescriptions for
defining Wick powers). It is simplest and most convenient to express
the formula for the difference between these prescriptions in terms of
the {\it local $S$-matrix}, $S(\sum f_i \Phi_i)$, for the formal sum $\sum f_i
\Phi_i$, which is formally defined by
\begin{equation}
\label{s}
S(\sum f_i \Phi_i) = \myid + \sum_{n \ge 1} \frac{i^n}{n!} T(\prod^n
\sum \Phi_i(f_i)).
\end{equation}
(Of course, as discussed further at the beginning of section 3.1
below, the series on the right side of eq.~(\ref{s}) is not expected
to converge. It should be viewed as merely a bookeeping device that
will allow us to write an infinite sequence of complicated
equations---given explicitly in eq.~(\ref{unique}) below---as a single
equation.)  Denote the image of the $n$-tuple $(f_1 \Phi_1, \dots,f_n
\Phi_n) \in \times^n \cD(M, \cV)$ under the first prescription as
$T(\prod \Phi_i(f_i))$ and denote its image under the second
prescription as $\tilde T(\prod \tilde \Phi_i(f_i))$. Then, if both
prescriptions satisfy all of the requirements stated in \cite{hw1},
\cite{hw2}, theorem~5.2 of \cite{hw1} establishes that the following
relation holds between the corresponding local $S$-matrices:
\begin{equation}
\label{SM}
\tilde S(\sum f_i \Phi_i) = S(\sum f_i \Phi_i + \delta(\sum f_i
\Phi_i)),
\end{equation}
where $\delta(\sum f_i\Phi_i)$ is given by the formal power series expression
\begin{equation}
\label{deldef}
\delta(\sum f_i\Phi_i) = \sum_{n \ge 1} \frac{i^{n-1}}{n!} 
O_n({\mbox{\huge $\times$}}^n \sum f_i\Phi_i).  
\end{equation}
Equation~(\ref{SM}) is to be interpreted as an infinite sequence of
equalities between terms containing equal numbers of each of the
$f_i$'s under the formal substitutions (\ref{s}) and
(\ref{deldef}). In eq.~(\ref{deldef}), the $O_n$'s are multilinear
maps
\begin{equation}
O_n: \mbox{\huge $\times$}^n \cD(M, \cV) \to \cD(M, \cV)
\end{equation}
of the form: 
\begin{eqnarray}
\label{on}
O_n(\times_{i=1}^n f_i \Phi_i) &=& \sum_{j} F_j \Psi_j,  
\end{eqnarray}  
where $\Psi_j$ are basis fields in $\cV$ and the densities $F_j$ 
are of the form
\begin{equation}
\label{Fa}
F_j(x) = \beps(x) \sum_{(a) = (a_1) \dots (a_n)} {C_{j}}^{(a)}(x)
\prod_{i=1}^n\nabla_{(a_i)} f_i(x) .
\end{equation}
In this formula, we have idenfied the densities $f_i$ with test functions on $M$
using the metric volume element $\beps$ and we have used the multi-index
notation $\nabla_{(a)} = \nabla_{(a_1} \cdots \nabla_{a_s)}$. The
quantities ${C_{j}}^{(a)}$ are tensors that are monomials in the
Riemann tensor, its covariant derivatives, and $m^2$, with coefficients that are 
analytic functions of $\xi$. The quantities $O_n$ are
further restricted by the requirement that 
\begin{equation}
\label{hdf}
[T(O_n(\times_{i=1}^n f_i \Phi_i)), \varphi(f_{n+1})] = \sum_{k=1}^n
T(O_n(f_1 \Phi_1, \dots, i\sum_{(a)} (f_{n+1} \Delta_{(a)} f_k)
\frac{\partial \Phi_k}{\partial \nabla_{(a)}\varphi}, \dots, f_n
\Phi_n)).
\end{equation}
Here, $\partial \Phi/\partial \nabla_{(a)} \varphi$ is the element in
$\cV$ obtained by formally differentiating the expression $\Phi \in
\cV$ with respect to $\nabla_{(a)} \varphi$ (thereby viewing the
latter as an ``independent variable''), $(a)$ is a spacetime
multi-index as above, and
\begin{equation}
\label{hdf1}
(f_{n+1} \Delta_{(a)} f_i)(x) = 
\int_M f_{n+1}(x)\Delta(x, y) \nabla_{(a)} f_i(y),
\end{equation} 
where $\Delta$ is the advanced minus retarded Green's function, and where
the integration is over the ``$y$''-variables. In addition, if
${{d_{j}}^{(a)}}$ is the engineering dimension of ${C_{j}}^{(a)}$,
$N_{(a)}$ the number of covariant derivatives appearing explicitly in
equation \eqref{Fa}, and $d_j$ is the engineering dimension of the field
$\Psi_j$, then each of the terms in the sum \eqref{Fa} must
satisfy the power counting relation
\begin{equation}
\label{pwct}
\sum_{i = 1}^n d_{\Phi_i} = 4n + N_{(a)} + {{d_{j}}^{(a)}} + d_{j} 
\end{equation}
for all multi-indices $(a)$ and all $j$. Furthermore, 
the quantities $\delta(f \Phi)$ defined in eq.~\eqref{deldef} satisfy the 
reality condition 
\begin{equation}
\label{realcon}
\delta(f \Phi)^* = \delta(f \Phi)
\end{equation}
for real valued $f$ and hermitian $\Phi$, 
which corresponds to the unitarity requirement, $S(f\Phi)^{-1} = S(f \Phi)^*$, 
for real valued $f$ and hermitian $\Phi$.
Equation~\eqref{realcon} is equivalent to the reality property 
$O_n( \times^n f\Phi)^* = (-1)^{n-1} O_n( \times^n f\Phi)$.

The relations between the two
prescriptions for time-ordered-products given implicitly in
eq.~(\ref{SM}) can be written out explicitly as
\begin{equation}
\label{unique}
\tilde T\left(\prod_{i=1}^n \tilde \Phi_{i}(f_i)\right) =  
T\left(\prod_{i=1}^n \Phi_{i}(f_i)\right) + 
\sum_{P} T\left(\prod_{I \in P} O_{|I|}
(\times_{j \in I} f_j \Phi_j) \prod_{i \notin I \,\, \forall I 
\in P} \Phi_{i}(f_i)\right).  
\end{equation}
where, $P$ is a collection of pairwise disjoint subsets $I_1, I_2,
\dots$ of the set $\{1, \dots, n\}$, not all of which can be empty,
and $|I|$ is the number of elements of such a
set. Equation~(\ref{unique}) corresponds to our previous
formulation of the uniqueness theorem given in theorem~5.2 of
\cite{hw1}, except that, for simplicity, we asssumed in the statement
of that theorem that the ``untilded'' prescription for defining
Wick products was given by ``local normal ordering'' with respect 
to a local Hadamard parametrix.  In Minkowski spacetime a proof that eq.~(\ref{unique})
corresponds to the formal expansion of eq.~(\ref{SM}) is given in
\cite[thm. 6.1]{p2}; the combinatorical arguments given there can be
generalized in a straightforward manner to the present case.

If we take $\sum f_i\Phi_i$ to be the interaction Lagrangian density, then 
eq.~(\ref{SM}) corresponds to the familiar statement in perturbative
quantum field theory in Minkowski spacetime that the ``renormalization
ambiguities'' in the $S$-matrix\footnote{We should emphasize that our
interest here is not in determining the renormalization ambiguities in
a global scattering matrix (which will, in general, not even be
defined) but rather in determining the renormalization ambiguities in
the interacting field itself (as well as its Wick powers and the
time-ordered-products of its Wick powers). However, the formulas
expressing these ambiguities are most conveniently expressed in terms
of the relative $S$-matrix, which is defined in terms of the local
$S$-matrix (see section 3.2 below), so a knowledge of the
ambiguities in the local $S$-matrix will enable us to determine the
ambiguities in the interacting field.} correspond simply to adding
``counterterms'' to the Lagrangian of the appropriate ``power
counting'' dimension. The only significant difference occurring when
one goes to curved spacetime is that additional counterterms involving
the spacetime curvature may occur.

We conclude this section by reviewing the scaling properties of Wick
powers and their time-ordered-products. Fix a Wick power $\Phi[M, \g, p]$ 
and consider the 1-parameter family of Wick powers
$\Phi[M, \lambda^2 \g, p(\lambda)]$ defined on the spacetimes $(M,  \lambda^2 \g)$, 
with coupling constants 
\begin{equation}
p(\lambda) =
(\lambda^{-2} m^2, \xi). 
\end{equation}
These quantities belong (when smeared with a
test density) to different algebras, 
\begin{equation}
\Phi[M, \lambda^2 \g, p(\lambda)](f) \in \cW(M, \lambda^2 \g,
p(\lambda))
\end{equation}
(where we now have indicated explicitly the dependence of
this algebra and the field on the coupling parameters $p$), and hence cannot be
compared directly. However, as observed in \cite{hw1}, one can
define a natural *-isomorphism
\begin{equation}
\label{bdef}
\sigma_\lambda: \cW(M, \lambda^2 \g, p(\lambda)) \to \cW(M, \g, p), 
\quad \sigma_\lambda ([t_n]) \equiv \lambda^{-n} [t_n].
\end{equation}
In other words, $\sigma_\lambda$ maps the element of 
$\cW(M, \lambda^2 \g, p(\lambda))$ corresponding to 
$\lno \varphi(f_1) \cdots \varphi(f_n) \rno_{\omega_\lambda}$ 
in the GNS-representation of the quasi-free Hadamard state 
$\omega_\lambda$ into the element of 
$\cW(M, \g, p)$ corresponding to 
$\lno \varphi(f_1) \cdots \varphi(f_n) \rno_{\omega}$ 
in the GNS-representation of the quasi-free Hadamard state $\omega$, where the 
two-point functions of $\omega_\lambda$ and $\omega$ are 
related by $\omega_\lambda(x_1, x_2) = \lambda^{-2} \omega(x_1, x_2)$.
Using this isomorphism, we can then identify the Wick product 
$\Phi[M, \lambda^2 \g, p(\lambda)]$ with a local covariant field 
$\sigma_\lambda (\Phi[M, \lambda^2 \g, p(\lambda)])$
for the unscaled metric and unscaled coupling constants $\g, p$.

The free field $\varphi$ has the homogeneous scaling behavior 
\begin{equation}
\sigma_\lambda (\varphi(f)) = 
\lambda^{-1} \varphi(f), 
\end{equation}
where the field on the left side of this equation is defined in terms of 
the scaled metric $\lambda^2 \g$ and scaled coupling constants $p(\lambda)$, 
whereas the field on the right side of this equation is defined in terms of 
the unscaled metric $\g$ and unscaled coupling constants $p$. The higher order Wick powers 
and their time-ordered-products
have an ``almost'' homogeneous scaling behavior in the sense that\footnote{The
fact that the non-homogeneous terms on the right side of
eq.~(\ref{freesc}) take the form of local, covariant fields that
depend polynomially on $\ln \lambda$ was taken as an axiom in
\cite{hw1}, the consistency of which was proven in \cite{hw2}. The
specific form of these terms follows from the uniqueness theorem of
\cite{hw1}.}
\begin{multline}
\label{freesc}
\sigma_\lambda \left( T \left( \prod_{i=1}^n {\Phi_i(f_i)} \right) \right)= \lambda^{-d_T} \,\,
T\left(\prod_{i=1}^n \Phi_{i}(f_i) \right) + \\
\lambda^{-d_T} \sum_{P} T \left(\prod_{I \in P} O_{|I|}
(\lambda; \times_{j \in I} f_j \Phi_j) \prod_{i \notin I \,\, \forall I 
\in P} \Phi_{i}(f_i) \right),
\end{multline}
where $d_T$ is the engineering dimension of the time-ordered-product
and the quantities 
\begin{equation}
O_n(\lambda; \times_{i=1}^n f_i \Phi_i) = 
\sum_j F_j(\ln \lambda) \Psi_j
\end{equation}
have the same properties as the quantities eq.~\eqref{Fa} in our uniqueness
theorem, with the only difference that the scalar densities $F_j(\ln \lambda)$
now have an additional {\it polynomial} dependence on $\ln \lambda$. 

As we will see, the fields in the interacting quantum field theory
will {\it not} have this almost homogeneous scaling behavior in general. 

\section{Interacting fields in curved spacetime}
\subsection{Definition of the interacting field}

In this section, we consider the interacting field theory
described by the Lagrangian density (\ref{L}). Our main aim is to
define the interacting field, $\varphi_{L_1}$, as well as its Wick
powers and the time-ordered-products of its Wick powers. We use the
generic notation $\Phi_{L_1}$ to denote any Wick power and $T_{L_1}(\prod \Phi_i)$
to denote any time-ordered-product of Wick powers of the interacting field.

The first step is to define a suitable algebra, $\cX(M,\g)$, of which
these interacting fields will be elements. The interacting
field algebra will then be defined to be a suitable subalgebra,
$\cB_{L_1}(M,\g)$, of $\cX(M,\g)$ (see eq.~(\ref{ifa})
below). Unfortunately, even in Minkowski spacetime, if $\kappa \neq 0$
there is no known way to construct the fields for this theory other
than on the level of perturbation theory. Furthermore, the
perturbative formulae for the quantities that are normally
calculated---such as Green's functions and $S$-matrix elements---are
not expected to converge. In this regard, however, we note that
quantities such as Green's functions and $S$-matrix elements do not
depend solely on the algebraic properties of the fields themselves,
but also involve properties of the vacuum state or ground state and,
in many instances, also ``in'' and ``out'' states. However, even if,
in some suitable sense, the algebra of 
fields were to vary analytically under
changes of the parameter $\kappa$, there is no reason that certain
states of the theory, such as the ground state, need vary
analytically. This suggests the possibility that if perturbation
theory were used solely for the purpose of calculating algebraic
relations involving the interacting field---rather than properties
involving states---then perhaps at least some of the difficulties with
the convergence of perturbative expansions would not arise. In other
words, rather than using perturbation theory to calculate Green's
functions, $S$-matrix elements, or other quantities that depend upon
states, we suggest that it may be more fruitful to use perturbation
theory to attempt to find analytic relations between the field
observables that hold to all orders in perturbation theory.

However, we shall not attempt to pursue any such program here, but
rather will only attempt to construct the interacting theory at the
level of formal power series in the coupling constant $\kappa$. Thus,
we shall take $\cX(M,\g)$ to be
\begin{equation}
\label{X}
\cX(M,\g) = \mbox{\huge $\times$}_{n=0}^\infty \cW(M,\g)
\end{equation}
where an element $A \in \cX(M,\g)$ of the form $A = (A_0, A_1, A_2,
\dots)$ should be interpreted as corresponding to the formal power series
\begin{equation}
\label{forpow}
A = \sum_{n=0}^\infty A_n \kappa^n.
\end{equation}
The multiplication law in $\cX(M,\g)$ is then defined to be that
corresponding to the multiplication of the formal power series
expressions (\ref{forpow}), i.e., if $A = (A_0, A_1, A_2, \dots)$ and
$B = (B_0, B_1, B_2, \dots)$, then $A \star B = (A_0 \star B_0, A_1
\star B_0 + A_0 \star B_1, \dots)$. Note that the interacting field
algebra $\cB_{L_1}(M,\g) \subset \cX(M,\g)$ that we will define in
eq.~(\ref{ifa}) below will then formally correspond to the entire one
parameter family of interacting field algebras for all values of
$\kappa$, rather than the interacting field algebra for a specific
value of $\kappa$.

To define the interacting field, we first consider a situation in
which the interaction is turned on only in some finite spacetime
region, i.e., we choose a cutoff function, $\theta$, of compact
support on $M$ which is equal to 1 on an open neighborhood of the
closure, $\bar{V}$, of some globally hyperbolic open region $V$ with
the property that $\Sigma \cap V$ is a Cauchy surface for $V$ for some
Cauchy surface $\Sigma$ in $M$. This cutoff will be removed in a later
step (see below).  We define the {\em relative $S$-matrix} for $f \Phi$
with respect to the interaction Lagrangian density $\theta \cL_1$ by
\begin{equation}
\label{rels}
\bS_{\theta L_1}(f\Phi) = S(\theta\cL_1)^{-1} \star S(\theta\cL_1 + f\Phi)
\end{equation}
where the local $S$-matrix, $S(f\Phi)$, was defined in eq.~(\ref{s}) above.
Then the Wick power, $\Phi_{\theta L_1}$, for the interacting theory
with Lagrangian density $\theta \cL_1$ corresponding to the Wick power $\Phi$ of the
free theory is defined by \cite{bs}
\begin{eqnarray}
\Phi_{\theta L_1}(f) &\equiv& 
\frac{\partial}{i\partial\alpha} \bS_{\theta L_1}(\alpha f \Phi) 
\bigg|_{\alpha=0}.  
\label{intfield}
\end{eqnarray}
Here the right side of eq.~(\ref{intfield}) should be viewed as
(rigorously) defining an element of $\cX(M,\g)$, which is obtained by
formally expanding $S(\theta\cL_1)^{-1}$ and $S(\theta\cL_1 + f\Phi)$
in powers of the coupling constant $\kappa$ and then collecting all of
the (finite number of) terms that multiply $\kappa^n$ for each $n$ (see
eq.~(\ref{forpow}) above and eq.~\eqref{retprod} below). Similarly, the time-ordered-product of Wick
powers of the interacting field with Lagrangian density $\theta \cL_1$ is
defined by
\begin{equation}
T_{\theta L_1} (\prod_{i=1}^n \Phi_i(f_i)) 
\equiv \frac{\partial^n}{i^n \partial \alpha_1 \dots \partial \alpha_n}
{\bS}_{\theta L_1}(\sum_i \alpha_i f_i \Phi_i) 
\bigg|_{\alpha_1 = \dots = \alpha_n = 0}. 
\label{inttop}
\end{equation} 

Note that the definition of $\Phi_{\theta L_1}$ (as well as that of
$T_{\theta L_1} (\prod \Phi_i)$) has been adjusted
so that $\Phi_{\theta L_1}$ coincides with the corresponding free
field $\Phi$ before the interaction is ``switched on''.
This can be seen explicitly by expressing  $\Phi_{\theta L_1}(f)$ 
in terms of the ``totally retarded products''\footnote{This formula is 
known as ``Haag's series,'' since an expansion of this kind was first derived 
in~\cite{haag} for Minkowski spacetime; see also~\cite{glz}.}   
\begin{equation}
\Phi_{\theta L_1}(f) = \Phi(f) + \sum_{n \ge 1} \frac{i^n}{n!}
R(f\Phi; \underbrace{\theta\cL_1, \dots, \theta\cL_1}_{n \,\, 
factors}),   
\label{retprod}
\end{equation}
Since the $R$-products have support
\begin{equation}
{\rm supp} R \subset \{(y, x_1, \dots, x_n) \mid x_i \in J^-(y) 
\quad \forall i \}, 
\end{equation}
it follows that all terms in the above sum will vanish if the support
of $f$ does not intersect the causal future of the support of
$\theta$.

Below, we will need to know how the fields \eqref{inttop} change under
a change of the cutoff function $\theta$. Now if $\theta$ and
$\theta'$ are two cutoff functions, each of which are 1 in an open
neighborhood of $\bar{V}$ as above, then there exists a smooth function $h_-$
of compact support on $M$ which is equal to $\theta - \theta'$ on the
causal past of the region $V$, and whose support does not intersect
the causal future of $V$. The unitary $U(\theta,\theta')$ defined by
\begin{equation}\label{Udef}
U(\theta,\theta') = \bS_{\theta  L_1}(h_- \cL_1) 
\end{equation}
is then independent of the particular choice for $h_-$, and
one has \cite[thm. 8.6]{bf1} 
\begin{equation}
\label{rel}
U(\theta, \theta') \star T_{\theta L_1}
(\prod \Phi_i(f_i)) \star U(\theta, \theta')^{-1} = 
T_{\theta' L_1}(\prod \Phi_i(f_i)),  
\end{equation}
for all fields $\Phi_i$ and all smooth scalar densities $f_i$
of compact support in $V$. 

We now remove the cutoff $\theta$. Formulas (\ref{intfield}) and
(\ref{inttop}) will not, in general, make sense if we
straightforwardly attempt to take the limit $\theta \to 1$. Indeed if
$\theta$ could be set equal to 1 throughout the spacetime in
eq.~(\ref{intfield}), then the resulting formula for $\Phi_{L_1}$
would define an interacting field in the sense of Bogoliubov
\cite{bs}, with the property that the interacting field approaches the
free field in the asymptotic past. However, even in Minkowski
spacetime, it is far from clear that such an asymptotic limit of the
interacting field will exist (particularly for massless fields), and it
is much less likely that any such limit would exist in generic
globally hyperbolic curved spacetimes that are not flat in the
asymptotic past.

In order to remove the cutoff in a manner in which the limit will
exist, we will not try to take a limit where the field remains fixed
in the asymptotic past but rather---following the ideas of
\cite{bf1}---we will take a limit where the field remains fixed in
regions of increasing size in the interior of the spacetime. To make
this construction precise, it is useful to have the following lemma:

\begin{lemma}
Let $(M, \g)$ be a globally hyperbolic spacetime. Then there exists a
sequence of compact sets, $\{K_n\}$, with the properties that (i) for
each $n$, $K_n \subset V_{n+1}$, where $V_{n+1} \equiv
{\rm int}(K_{n+1})$ (ii) $\cup_n K_n = M$, and (iii) for each $n$,
$V_n$ is globally hyperbolic and $\Sigma \cap V_n$ is a Cauchy surface
for $V_n$, where $\Sigma$ is a Cauchy surface for $M$.
\end{lemma}
\begin{proof}
Let $t$ be a time function on $(M, \g)$ with range $-\infty < t <
\infty$ whose level surfaces are Cauchy surfaces, $\Sigma_t$, that
foliate $M$ \cite{ger}, \cite{die}. Let $\Sigma = \Sigma_0$. Choose
any complete Riemannian metric, $q_{ab}$, on $\Sigma$, choose $x_0 \in
\Sigma$, and let $B_n$ be the closed ball (on $\Sigma$) of radius $n$
about $x_0$ with respect to $q_{ab}$. Define
\begin{equation}
K_n = D(B_n) \cap J^-(\Sigma_n) \cap J^+(\Sigma_{-n})
\label{Kn}
\end{equation}
where $D$ denotes the domain of dependence and $J^-$ and $J^+$ denote
the causal past and future, respectively. Then $K_n$ is
closed. Furthermore, since $B_n$ is compact it follows that $J^+(B_n)
\cap J^-(\Sigma_n)$ and $J^-(B_n) \cap J^+(\Sigma_{-n})$ are
compact. Since $K_n$ is a subset of the union of these two sets, it
follows that $K_n$ is compact. Clearly, we have $V_n \subset
V_{n+1}$. However, if $x$ lies on the boundary of $K_n$,
then it must lie on the boundary of $D(B_n)$ and/or lie on $\Sigma_n$
or $\Sigma_{-n}$; in all cases, it follows immediately that $x \in
V_{n+1}$. Thus, $K_n \subset V_{n+1}$. To prove property (ii), let $y \in M$ with, say, $y \in
J^+(\Sigma)$. Since $J^-(y) \cap \Sigma$ is compact, it must be
contained in some ball of radius $r$ about $x_0$ (with respect to the
metric $q_{ab}$ on $\Sigma$). Then $y \in D(B_r)$, so $y \in K_n$ for
any $n$ such that $n > r$ and $n > t(y)$, as we desired to
show. Finally, the fact that $V_n$ is globally hyperbolic with Cauchy
surface $V_n \cap \Sigma$ follows immediately from the fact that $V_n$
is the interior of the domain of dependence of $B_n$ for the spacetime
$I^-(\Sigma_n) \cap I^+(\Sigma_{-n})$.
\end{proof}

Let $\{K_n\}$, $n = 1,2,\dots$, be a sequence of compact sets with the
properties stated in lemma 3.1. For each $n$, let $\theta_n$ be a
smooth function with support contained in $K_{n+1}$ such that
$\theta_n = 1$ on an open neighborhood of $K_n$. Let $U_1 = \myid$ and let
$U_n = U(\theta_n, \theta_{n-1})$ for all $n>1$, where $U(\theta_n, 
\theta_{n-1})$ was defined in eq.~(\ref{Udef}) above. Write $u_n = U_1 \star
U_2 \star \dots \star U_n$. Our definition of the interacting field, its Wick
powers, and their time-ordered-products is:
\begin{equation}
T_{L_1}(\prod \Phi_i(f_i)) \equiv \lim_{n \rightarrow \infty}
{\rm Ad} (u_n) \, T_{\theta_n L_1}(\prod
\Phi_i(f_i)),   
\label{intflddef}
\end{equation}
where we use the notation ${\rm Ad}(u_n)A = u_n \star A \star u_n^{-1}$ for any $A \in \cX(M, \g)$.
The existence of the limit is a direct consequence of the following
proposition:

\begin{prop}
Suppose that $N$ is such that the support of each $f_i$ is contained
in $K_N$. Then for all $n,m \geq N$ we have
\begin{equation}
{\rm Ad}(u_n) \,T_{\theta_n L_1}(\prod \Phi_i(f_i)) = {\rm Ad}(u_m) \, T_{\theta_m L_1}(\prod \Phi_i(f_i))
\label{nm}
\end{equation}
\end{prop}
\begin{proof}
It suffices to show that for any $n \geq N$ we have 
\begin{equation}
u_{n+1} \star T_{\theta_{n+1} L_1}(\prod \Phi_i(f_i)) \star
u^{-1}_{n+1} = u_n \star T_{\theta_n L_1}(\prod \Phi_i(f_i))
\star u^{-1}_n
\label{nm2}
\end{equation}
But by eq.~(\ref{rel}) we have
\begin{equation}
U_{n+1} \star T_{\theta_{n+1} L_1}(\prod \Phi_i(f_i)) \star
U^{-1}_{n+1} = T_{\theta_n L_1}(\prod \Phi_i(f_i))
\label{nm3}
\end{equation}
from which the desired result follows immediately by applying 
${\rm Ad}(u_n)$ to both sides.
\end{proof}

Now, given any compact set $K \subset M$ and any family of compact
sets $K_n$ satisfying properties (i) and (ii) of the above lemma, then
there always exists\footnote{Proof: Otherwise, one could find a
sequence $\{x_n\} \in K$ such that $x_n \notin K_n$ for all
$n$. However, this sequence would have an accumulation point, $x$,
which must lie in the interior of some $K_N$, resulting in a
contradiction.} an $N$ such that $K \subset K_N$. Given any smeared
time-ordered-product of Wick powers, we choose $K$ to be the union of
the supports of all of the (finite number of) test functions appearing
in the time-ordered product. By the above proposition, there exists an
$N$ such that the sequence appearing on the right side of
eq.~(\ref{intflddef}) is constant for all $n > N$. Therefore, the
limit exists.

The meaning of the sequence ${\rm Ad}(u_n) \,T_{\theta_n L_1}(\prod
\Phi_i(f_i)), n = 1, 2, \dots$, is easily understood as follows. Since
$u_1 = \myid$, the first element of this sequence is just the
Bogoluibov formula for this interacting field quantity with cutoff
function $\theta_1$. The second element of this sequence modifies the
Bogoliubov formula with cutoff function $\theta_2$ in such a way that,
according to eq.~(\ref{rel}) above, the modified Bogoliubov formula
with cutoff function $\theta_2$ agrees with the unmodified Bogoliubov
formula with cutoff function $\theta_1$ when the supports of all of
the $f_i$ are contained within $K_1$. For the third element of the
sequence, the unitary map $U_3$ first modifies the Bogoliubov formula
with cutoff function $\theta_3$ so that it agrees in region $K_2$ with
the Bogoliubov formula with cutoff function $\theta_2$. The action of
the unitary $U_2$ then further modifies this expression so that it
agrees in region $K_2$ with the modified Bogoliubov formula of the
previous step. In this way, we have implemented the idea of ``keeping
the interacting field fixed in the interior of the spacetime'' as the
cutoff is removed.

We define the {\em interacting field algebra} $\cB_{L_1}(M, \g)$ to be
the subalgebra of ${\mathcal X} (M, \g)$ generated by the interacting
field, its Wick powers, and their time-ordered-products, i.e.,
\begin{equation}
\cB_{L_1}(M, \g) \equiv \{ \text{algebra generated by 
$T_{L_1}(\prod \Phi_i(f_i))$} 
\mid f_i \in \cD(M), \Phi_i \in \cV \}.
\label{ifa}
\end{equation}
This definition of $\cB_{L_1}(M, \g)$ as a subalgebra of ${\mathcal X} (M,
\g)$ depends on a choice of a family of compact sets $K_n$ satisfying
the properties of lemma 3.1 as well as a choice of cutoff functions
$\theta_n$. If we were to choose a different family, $\tilde{K}_n$, of
compact sets and a corresponding different family, $\tilde{\theta}_n$,
of cutoff functions, we will obtain a different subalgebra
$\tilde{\cB}_{L_1}(M, \g) \subset {\mathcal X} (M, \g)$ of interacting
fields. However, the algebra $\tilde{\cB}_{L_1}(M, \g)$ is isomorphic
to $\cB_{L_1}(M, \g)$. To see this, focus attention on the subalgebras
$\tilde{\cB}_{L_1}(K, \g)$ and $\cB_{L_1}(K, \g)$ generated by
fields that are smeared with test functions with support in a fixed
compact set $K$. Let $n$ be such that $K \subset K_n$ and $K
\subset \tilde{K}_n$. Let 
\begin{equation}
X_n =  u_n \star U(\tilde{\theta}_n, 
\theta_n) \star \tilde{u}^{-1}_n. 
\end{equation}
Then $X_n$ is a unitary element of
${\mathcal X} (M, \g)$. However, for any $\tilde{F} \in
\tilde{\cB}_{L_1}(K, \g)$, it follows from eqs.~(\ref{rel}) and
(\ref{intflddef}) together with proposition 3.1 that ${\rm Ad}(X_n) \tilde{F}$ 
is the corresponding interacting field quantity $F \in
\cB_{L_1}(K, \g)$. This shows that the map $\gamma_K :
\tilde{\cB}_{L_1}(K, \g) \rightarrow \cB_{L_1}(K, \g)$ which
associates to any element of $\tilde{\cB}_{L_1}(K, \g)$ the
corresponding interacting field quantity in $\cB_{L_1}(K, \g)$ is
well defined and is a *-isomorphism. However, since $K$ is arbitrary,
this argument actually shows that the map $\gamma : \tilde{\cB}_{L_1}(M,
\g) \rightarrow \cB_{L_1}(M, \g)$ which associates to any element of
$\tilde{\cB}_{L_1}(M, \g)$ the corresponding element of $\cB_{L_1}(M,
\g)$ also is well defined and is a *-isomorphism of these
algebras\footnote{Note, however, that there need not
exist a unitary element $X \in {\mathcal X} (M, \g)$ whose action on
$\tilde{\cB}_{L_1}(M, \g)$ coincides with $\gamma$.}. Thus, as an
abstract algebra, $\cB_{L_1}(M, \g)$ is independent of the choices of
$K_n$ and $\theta_n$ that entered in its construction.  In the
following we assume that we have made an arbitrary, but fixed, choice
for $K_n$ and $\theta_n$ in every spacetime.

In the free theory, the notion of a local and covariant field was
defined relative to a natural injective *-homomorphism $\alpha_\chi:
\cW(\tilde M, \tilde \g) \to \cW(M, \g)$ associated with causality
preserving isometric embeddings $\chi$ of a spacetime $(\tilde M,
\tilde \g)$ into another spacetime $(M, \g)$. The Wick products of the
free field and their time-ordered-products were then seen to be local,
covariant fields in the sense that eq.~(\ref{lcf}) holds. In order to
get a corresponding natural injective *-homomorphism,
${\mbox{\boldmath $\alpha$}}_\chi:\cB_{L_1}(\tilde M, \tilde \g) \rightarrow
\cB_{L_1}(M,\g)$, for the interacting field algebra, we must compose the
natural action of $\alpha_\chi$ on $\cB_{L_1}(\tilde M, \tilde \g)$ with
the map $\gamma$ constructed above in order to compensate for the fact
that the choices for $K_n$ and $\theta_n$ on $(M, \g)$ may not
correspond to the choices of $\tilde{K}_n$ and $\tilde{\theta}_n$ on
$(\tilde M, \tilde \g)$. It then follows that the interacting field,
its Wick powers and their time-ordered-products as defined above are
local and covariant fields in the sense that for any causality
preserving isometric embedding, $\chi$, we have
\begin{equation}
\label{intlcf}
{\mbox{\boldmath $\alpha$}}_\chi(\Phi_{L_1}[\tilde M, \tilde \g](x)) = 
\Phi_{L_1}[M, \g](\chi(x)), 
\end{equation}
with an analogous equation holding for the interacting time-ordered-products.

Finally, we comment upon how the theory we have just
defined is to be interpreted, i.e., how the mathematical formulas
derived above for the interacting field relate to predictions of
physically observable phenomena. In many discussions of quantum field
theory in Minkowski spacetime, the interpretation of the theory is
made entirely via the (global) $S$-matrix. Here it is assumed that in
the asymptotic past and future, states of the field can be identified
with states of a free field theory, which have a natural particle
interpretation. It is also assumed that one can prepare states
corresponding to desired incoming particle states and that one can
measure the properties of the state of outgoing particles, so that the
$S$-matrix can be determined. A wide class of predictions of the
theory---including essentially all of the ones that can be measured in
practice---can thereby be formulated in terms of measurements of the
$S$-matrix for particle scattering, without the need to even mention
local fields. Indeed, when this viewpoint on quantum field theory is
taken to the extreme, the local quantum fields, in effect, play the
role of merely being tools used for calculating the $S$-matrix.

An alternative, but closely related, viewpoint on interpreting the
theory in Minkowski spacetime makes crucial use of the existence of a
preferred vacuum state. Here, one focuses attention on the correlation
functions of the field in this state, which are assumed to be
measureable---at least in the asymptotic past and future and for
sufficiently large spatial separation of the points. The
interpretation of the theory can be formulated in terms of its
predictions for these correlation functions. This viewpoint on the
interpretation of the theory is closely related to the first one,
since the particle measurements in the $S$-matrix interpretation can
be viewed as really corresponding to measuring certain properties of
these correlation functions.

However, for quantum fields in a general, globally hyperbolic curved
spacetime, we do not expect to have asymptotic, free particle states
or any globally preferred states. It therefore would not appear
fruitful to attempt to interpret the theory in a manner analogous to
the above ways in which the theory is normally interpreted in
Minkowski spacetime. Rather, it would seem much more fruitful to view
the interacting field itself---together with its Wick powers and other
local covariant fields in $\cB_{L_1}(M, \g)$---as the fundamental
observables in the theory. To make ``measurements'', we assume that we
have access to some external systems that couple to the field
observables of interest via known interaction Lagrangians, and that
we can then measure the state of the external systems at different
times. It is clear that by making sufficiently many measurements of
this sort, we can test any aspect of the theory and---if the theory is
valid---we also can determine any unknown coupling parameters in the
theory. However, it is not straightforward to give a simple, universal
algorithm for doing so, since the properties of the states will depend
upon the spacetime under consideration, and a type of experiment that
would most usefully probe the theory for a particular spacetime may
not be as useful for another spacetime.

To make the remarks of the previous paragraph more explicit, consider
a typical experiment in Minkowski spacetime wherein one prepares a
system of particles in a given incoming state and measures the
particle content of the outgoing particles. Both the ``state
preparation'' and the ``measurement'' of the ``particles'' in their
final state really consist of introducing certain external systems
that have desired couplings to the quantum field, preparing the
initial state of these external systems suitably, and measuring their
final state. In a curved spacetime, one could presumably introduce
external systems with couplings to the field that are similar to those
of systems used in Minkowski spacetime, although it should be noted
that there is not any obvious, general notion of what it means to have ``the
same'' system in a curved spacetime as one had in Minkowski spacetime,
unless one goes to a limit where the size of the system is much
smaller than any curvature scales. However, even if one considers an
external system in curved spacetime that corresponds to a system of
``particle detectors'' in Minkowski spacetime, it may not be possible
to give any consistent interpretation of the outcome of the curved
spacetime measurements in terms of ``particles''. Nevertheless, such
measurements still provide information about the states of the quantum
field, and it is clear that all aspects of the quantum field theory
can be probed by coupling the field to suitable external systems and
measuring the state of these external systems.

In should be noted that the above situation is not significantly
different from the case of classical field theory. Suppose that a
classical field $\varphi$ with Lagrangian (\ref{L}) can be measured
via its effect on the motion of scalar test charges, which feel a
force proportional to $\nabla_a \varphi$. In Minkowski spacetime, one
could set up an experiment where a global family of inertial observers
release test particles at some time in the distant past. By studying
the test particle motion for a brief interval of time, they could
reconstruct $\varphi$ (up to a constant) in that region of spacetime
and associate a noninteracting solution with the state of the field in
the distant past. By repeating this procedure in the distant future
they could obtain a corresponding non-interacting solution there, and
they could thereby determine the classical $S$-matrix. A great deal of
information about the interacting theory is encoded in the classical
$S$-matrix. However, it does not seem straightforward to give a simple
algorithm for making measurements with a similar interpretative
content in a general curved spacetime, where there are no asymptotic
regions and no globally preferred families of observers. Nevertheless,
it is clear that the classical field theory in curved spacetime is as
meaningful and interpretable as in Minkowski spacetime, and that all
of the predictions of the curved spacetime theory can be probed by
doing experiments that study the motion of a sufficiently wide class
of test particles.

\subsection{Renormalization ambiguities for the interacting 
field}

In the previous subsection we explained the construction of the
interacting Wick products and their time-ordered-products in the
interacting field theory classically described by the Lagrangian $\cL$
given by \eqref{L}. These constructions were based on a prescription
for defining the Wick products and their time-ordered-products in the
corresponding free field theory. As we discussed in section 2, the
definition of these quantities is subject to some well-specified
ambiguities. Therefore, the quantities in the interacting field
theory also will be subject to ambiguities.  

The purpose of this section is to give a precise specification of
these ambiguities. We shall show is that a change in the prescription
for the Wick products and their time-ordered-products (within the
class of ``allowed prescriptions'' specified by our uniqueness
theorem) corresponds to a shift of coupling parameters of the theory
appearing in the Lagrangian \eqref{L}. More precisely, the interacting
field algebra obtained with the new prescription will be isomorphic to
the interacting field algebra obtained with the original prescription,
but with the interaction Lagrangian modified by the addition of
``counterterms'', which---for a renormalizable theory, as considered
here---are of the same form as those appearing in the original
Lagrangian. This isomorphism of the interacting field algebras for the
two different prescriptions will map the interacting field to a
multiple of the interacting field. However, the relationship between
the higher Wick powers of the interacting field and their
time-ordered-products as defined by the two prescriptions is more
complicated: the isomorphism between the algebras will map a higher
Wick power (or a time-ordered-product of Wick powers) into a field of
the form specified in eq.~(\ref{VAV}) below.

To make the above statements more explicit, suppose that we are given
two prescriptions for defining the Wick products and their time
ordered products in the free field theory, both satisfying the
assumptions of our uniqueness theorem. These prescriptions will give
rise to two different constructions of interacting fields, which we
shall denote as $T_{L_1}(\prod \Phi_i)$ respectively $\tilde
T_{L_1}(\prod \tilde \Phi_i)$, and we write 
$\cB_{L_1}(M, \g)$ respectively $\tilde \cB_{L_1}(M, \g)$ for the 
algebras generated by these fields. Then the
relation between the tilde interacting fields and the untilde
interacting fields can be stated as follows: There exists a
*-isomorphism
\begin{equation}
\j: \tilde \cB_{L_1}(M, \g) \to \cB_{L_1 + \delta L_1}(M, \g)
\end{equation}
such that 
\begin{equation}
\label{VAV0}
\j \big( \tilde \varphi_{L_1}(f) \big) = Z \varphi_{L_1 + \delta L_1}(f),
\end{equation}
for all $f \in \cD(M)$.  The field $\tilde \varphi_{L_1}$ on the left
side of eq.~\eqref{VAV0} is the interacting field defined using the
``tilde prescription'' with respect to the interaction Lagrangian density
$\cL_1$, whereas the field $\varphi_{L_1 + \delta L_1}$ on the right
side of this equation is defined using the ``untilde prescription''
with respect to the interaction Lagrangian density $\cL_1 + \delta
\cL_1$, where $\delta \cL_1$ is given by
\begin{equation}
\label{countert}
\delta \cL_1 = \frac{1}{2}[\delta z (\nabla\varphi)^2 + 
\delta \xi R \varphi^2 + \delta m^2 \varphi^2 + \delta \kappa \varphi^4]
\beps.
\end{equation}
The parameters in this expression (including $\delta \kappa$), as 
well the parameter $Z$ in eq.~\eqref{VAV0} are formal
power series in $\kappa$ with real coefficients.  The generalization
of formula~\eqref{VAV0} for the action of $\j$
on an arbitrary interacting time-ordered-product in the
tilde prescription is given by
\begin{multline}
\label{VAV}
\j \left( \tilde T_{L_1}\left(\prod_{i=1}^n \tilde \Phi_i (f_i) \right) \right)  = 
T_{L_1 + \delta L_1} \left(\prod_{i=1}^n Z_i\Phi_i (f_i) \right)+ \\
\sum_{P} T_{L_1 + \delta L_1} \left(\prod_{I \in P} \bO_{|I|}(\times_{i \in I} f_i
\Phi_i) 
\prod_{j \notin I \, \forall I \in P} Z_j \Phi_j(f_j) \right).      
\end{multline}
Here, the $Z_i$ are formal power series in $\kappa$ whose coefficients are real 
provided the corresponding field $\Phi_i$ is (formally) hermitian.  
The $\bO_n$ are multilinear maps from $\times^n \cD(M, \cV) 
\to \cD(M, \cV)$ that depend on the interaction Lagrangian $\cL_1$ and 
have similar  properties to the maps $O_n$ in our uniqueness theorem for the 
time-ordered products of Wick products in the free theory:
First, the $\bO_n$ can be given an
analogous representation to the quantities $O_n$ 
in the free theory given in eq.~\eqref{on}, 
\begin{equation}
\label{on1}
\bO_{n}( \times_{i=1}^n f_i \Phi_i) = 
\sum_j c_j G_j \Psi_j
\end{equation}
The densities $G_j$ have the same form as the 
the corresponding expressions $F_j$ in the free theory 
(see eq.~\eqref{Fa}), and the $c_j$ are  
formal power series in $\kappa$.
If the terms appearing on the 
right side of eq.~\eqref{on1} are written out in terms of 
geometrical tensors (and the coupling constants in the free theory), 
then the engineering dimensions of 
each term will satisfy a ``power counting relation'' identical to 
that in the free theory, eq.~\eqref{pwct}. 

In terms of the generating functional
\begin{equation}
\label{genf}
\bS_{L_1}(\sum f_i\Phi_i) = 
\myid + \sum_{n\ge 1} \frac{i^n}{n!} T_{L_1}(\prod^n \sum \Phi_i(f_i))
\end{equation}
for the interacting Wick products and time-ordered-products,  
and the generating functional
\begin{equation}
\label{bdel}
\bdel_{L_1}(\sum f_i \Phi_i) \equiv \sum_{n \ge 1} \frac{i^{n-1}}{n!} 
\bO_n({\mbox{\huge $\times$}}^n \sum f_i\Phi_i), 
\end{equation}
relations~\eqref{VAV} can be rewritten more compactly as 
\begin{equation}
\j \left( \tbS_{L_1}(\sum f_i \Phi_i) \right) =  
\bS_{L_1 + \delta L_1}(\sum Z_i f_i \Phi_i + \bdel_{L_1}(\sum f_i \Phi_i)).
\end{equation}

In the preceding discussion, we have highlighted the analogies between
the structure of the renormalization ambiguities in the free and
interacting theories. However, there are also some key differences.
Firstly, in our identity~\eqref{unique} specifying the renormalization
ambiguities of the time-ordered-products in the free theory, the tilde
and untilde time-ordered-products are defined both ``with respect to
the same Lagrangian''.  By contrast, in the corresponding
formula~\eqref{VAV} in the interacting theory, the tilde and untilde
time-ordered-products are defined with respect to different
Lagrangians.  A second key difference between formulas~\eqref{unique}
and~\eqref{VAV} the free and interacting theories is the
appearance of the ``field strength renormalization factors,'' $Z_i$,
in the interacting theory, which are absent in the free theory.
Third, while the maps $O_n$ and $\bO_n$ in the free and interacting
theories satisfy a number of similar properities, the map $\bO_n$ does
not satisfy the commutator property, eq.~\eqref{hdf}, satisfied by
$O_n$ in the free theory.  Fourth, we note the appearance of the
automorphism $\j$ in our formula~\eqref{VAV} for the renormalization
ambiguity of the interacting time-ordered-products, which is absent in
the corresponding formula~\eqref{unique} in the free theory.

\vspace{0.5cm}
\noindent
\paragraph{Proof of equation \eqref{VAV}:}
Let $\theta$ be a cutoff function of compact support as above which is
1 in an open neighborhood of the closure, $\bar{V}$ of a globally
hyperbolic subset $V$ of $M$ such
that $V \cap \Sigma$ is a Cauchy surface of $V$ for some
Cauchy surface $\Sigma$ of $M$.  Eq.~\eqref{SM} implies that
\begin{equation}
\label{SM'}
\tbS_{\theta L_1}(f \Phi) = S(\theta \cL_1 + \delta(\theta \cL_1))^{-1} \star
S(f\Phi + \theta \cL_1 + \delta(f\Phi + \theta \cL_1)). 
\end{equation} 
In order to bring this equation into a more convenient form, let us define the 
following elements in ${\mathcal X}(M, \g)$: 
\begin{equation}
\delta_n(\theta \cL_1; f_1 \Phi_1, \dots, f_n \Phi_n) \equiv
\frac{\partial^n}{i^{n-1} \partial \alpha_1 \dots \partial \alpha_n}
\delta(\theta \cL_1 + \sum_{i = 1}^n \alpha_i f_i \Phi_i) 
\bigg|_{\alpha_1 = \dots
= \alpha_n = 0}. 
\end{equation}
It follows from our uniqueness theorem that we can write $\delta_0(\theta \cL_1)$ as 
a sum (over $n$ and $j$) of terms of the general form 
\begin{equation}
\label{Faa}
F_{n,j}(x) \Psi_j = \beps(x) \sum_{(a) = (a_1) \dots (a_n)} {C_{n,j}}^{(a)}(x)
\prod_{i=1}^n \nabla_{(a_i)} \theta (x) \Psi_j,
\end{equation}
where ${C^{(a)}}_{n,j}$ are monomials in the Riemann tensor, its derivatives, and 
$m^2$. Since $\theta \cL_1$ has engineering dimension 4, it follows from eq.~\eqref{pwct}
that each term in~\eqref{Faa} must have engineering dimension 4. 
Since $\theta \cL_1$ is hermitian, it follows from eq.~\eqref{realcon} that the $C^{(a)}_{n,j}$
must be real and that the fields $\Psi_j$ must be hermitian. 
We now divide the terms~\eqref{Faa} appearing in $\delta_0(\theta \cL_1)$ into a
group consisting of all terms not containing any derivatives of $\theta$ and a
second group of terms each containing at least one derivative of $\theta$. This gives a 
decomposition of $\delta_0(\theta \cL_1)$ into the following two groups of terms:
\begin{eqnarray}
\label{decomposition}
\delta_0(\theta \cL_1) = \beps \sum_{n \ge 1} \kappa^n \theta^n \sum_j c_{n, j} \Psi_j  
+ \sum_{n \ge 1} \kappa^n \sum_j f_{n, j} \bLa_j.    
\end{eqnarray}
Here, $c_{n, j}$ are real constants, $\Psi_j$ runs through all hermitian fields of engineering dimension 4
(including fields with dimensionful couplings such as $m^2 \varphi^2$ or $R^2 \myid$), the $f_{n,j}$ 
are compactly supported smooth densities on $M$ whose support does not intersect on open neighborhood of $\bar V$, and 
$\bLa_j$ are hermitian fields of engineering dimension less than 4. In the decomposition~\eqref{decomposition}, 
we may replace the smooth functions $\theta^n$ in the first sum by the function $\theta$ at the expense of 
adding new terms of the kind appearing in the second sum, except that these new terms will have 
engineering dimension equal to 4. If this is done, we obtain the decomposition
\begin{equation}
\label{deco}
\delta_0(\theta \cL_0) = \theta \delta \cL_1 + \sum_j h_j \bLa_j.
\end{equation}
Here $\delta \cL_1$ is the real linear combination 
$\beps \sum a_j \Psi_j$ where $\Psi_j$ is running over 
all hermitian fields of engineering dimension 4 (including again fields with dimensionful coupling) 
and where $a_j = \sum_{n\ge 1} c_{n,j} \kappa^n$. 
The second sum in the above decomposition~\eqref{deco} of $\delta_0(\theta \cL_1)$
contains only real test densities $h_j$ of compact support that vanish
on an open neighborhood of $\bar{V}$.  The quantities $\bLa_j$ 
are now hermitian fields of engineering dimension $\le 4$. 

The field (density) $\delta \cL_1$ in eq.~\eqref{deco} 
is therefore of the form claimed in eq.~\eqref{countert}, except that it may 
contain (i) terms of the form $C_j \myid$, where $C_j$ is a monomial in the Riemann tensor, its 
covariant derivatives and 
$m^2$, and (ii) a term proportional to $\varphi \square \varphi$. In principle these terms should 
be included in eq.~\eqref{countert}. However, 
the terms (i) proportional to the identity do not contribute to the relative $S$-matrix given 
by eq.~\eqref{SM'} and can therefore be dropped. Furthermore, it can be seen that the term 
(ii) can always be eliminated in favor of the term $m^2 \varphi^2 + \xi R \varphi^2$
together with a sum of products of curvature tensors and $m^2$ of
engineering dimension 4 times the identity $\myid$,  
if the following additional condition is imposed on the time-ordered-products:
\begin{equation}
\label{63}
T\left( \varphi(\square - m^2 - \xi R) \varphi(f_0) \prod_{i=1}^n \Phi_i(f_i) \right) = 
T\left( \sum_j K_j \myid (f_0) \prod_{i=1}^n \Phi_i(f_i) \right)
\end{equation}
for all $\Phi_i$ and all $f_i \in \cD(M)$, where $K_j$ are monomials in the Riemann tensor, its 
derivatives and $m^2$ of engineering dimension 4. For the case of the Wick power 
$\varphi(\square - m^2 - \xi R)\varphi$ itself, this condition was 
shown to hold by Moretti~\cite[eq. (47)]{mor} for the ``local normal ordering prescription'' 
given in~\cite{hw1} and eq.~\eqref{lwprod} below. Using the methods of~\cite{hw2}, it can be shown that this 
additional normalization condition can also be satisfied for general time-ordered-products of the 
form~\eqref{63}. 
Therefore, we will assume that a condition of the form eq.~\eqref{63}
has been imposed\footnote{We will give a systematic analysis elsewhere 
of conditions that can be imposed on Wick powers and time-ordered-products involving derivatives.}. 
It then follows that $\delta \cL_1$ has the form claimed in eq.~\eqref{countert}.

Again, using the properties of the maps $O_n$ in our
uniqueness theorem, we can write
\begin{equation}
\delta_1(\theta \cL_1; f\Phi) = f \delta Z \Phi + \bO_1(f\Phi),   
\end{equation}
where $\delta Z$ is a formal power series in the coupling constant 
$\kappa$. If $\Phi$ is hermitian, then it follows again from eq.~\eqref{realcon} 
that these power series have real coefficients. The element  
$\bO_1(f\Phi)$ is of the form $\sum Z_j G_j \Psi_j$, 
where the $G_j$ can be written as
\begin{equation}
G_j(x) = \beps(x) \sum_{(a)} {C_j}^{(a)}(x) \nabla_{(a)} f(x), 
\end{equation}
where we have identified the density $f$ with a smooth function on $M$ via the 
metric volume element $\beps$ and where the ${C_j}^{(a)}$ are monomials in the Riemann tensor, 
its derivatives and $m^2$ of the correct dimension. The $Z_j$ are formal power series in $\kappa$ and 
the $\Psi_j$ are local covariant fields with fewer powers in the 
free field than $\Phi$. Moreover, for $n \ge 2$, we define 
\begin{equation}
\bO_n(\times_{i=1}^n f_i \Phi_i) \equiv 
\delta_n(\theta \cL_1; f_1 \Phi_1, \dots
, f_n \Phi_n).
\end{equation}   
Using the properties of $O_n$ given in our uniqueness theorem for the
time-ordered-products in the free theory, we can again conclude that
the $\bO_n$ must have the form stated below eq.~\eqref{VAV}, and that, 
in particular, they are independent of the particular choice of
$\theta$ so long as the support of $f$ is contained in the region
where $\theta$ is equal to 1. If we finally define $\bdel_{\theta
L_1}(f\Phi)$ as in eq.~\eqref{bdel} and set $Z = 1 + \delta Z$, then
we can recast eq.~\eqref{SM'} into the following form:
\begin{equation}
\label{start}
\tbS_{\theta L_1}(f \Phi) =
\bS_{\theta(L_1 + \delta L_1) + \sum h_j \Lambda_j}(Z f \Phi + 
\bdel_{L_1}(f \Phi)). 
\end{equation}
On $J(V) = J^+(V) \cup J^-(V)$ (the union of causal future and causal
past of $V$), we decompose $h_j = h_{j-} + h_{j+}$, where
$h_{j\pm}$ has compact support which does not intersect $J^\mp(V)$. If
we now set
\begin{equation}
\label{Wdef}
W(\theta) = \bS_{\theta(L_1 + \delta L_1)}(\sum h_{j-} \bLa_j). 
\end{equation}
then we obtain by \cite[thm. 8.1]{bf1}, 
\begin{equation}
\label{swsw}
\tbS_{\theta L_1}(f \Phi) 
= W(\theta) \star
\bS_{\theta(L_1 + \delta L_1)} (Z f \Phi + \bdel_{L_1}(f\Phi)) \star W(\theta)^{-1},  
\end{equation}
which holds for all $f\in \cD(M)$ with compact support in $V$. More generally, an 
analogous formula will hold if the expression $f\Phi$ is replaced 
by a sum of the form $\sum \alpha_i f_i \Phi_i$, where each 
$f_i$ has compact support in $V$. 

We now obtain the desired formula eq.~\eqref{VAV} from
eq.~\eqref{swsw} by removing the cutoff $\theta$ in the same 
way as in our definition of the interacting field in section~3.1:
We consider a sequence of cutoff functions $\theta_n$ which are
equal to 1 on globally hyperbolic open regions $V_n$ with compact 
closure that exhaust $M$. The interacting fields $T_{L_1 + \delta L_1}(\prod
\Phi_i)$ are then given in terms of the corresponding fields with cutoff
interaction $\theta_n(\cL_1 + \delta \cL_1)$ via
eq.~\eqref{intflddef}, and the interacting fields $\tilde
T_{L_1}(\prod \tilde \Phi_i)$ are likewise given in terms of the corresponding
fields with cutoff interaction $\theta_n \cL_1$ by the tilde version
of eq.~\eqref{intflddef}. Using that the interacting fields with cutoff $\theta_n$
are related via the unitary 
$W(\theta_n)$ (see eq.~\eqref{swsw}), one can now
easily obtain a *-isomorphism $r: \tilde \cB_{L_1}(M, \g) \to 
\cB_{L_1 + \delta L_1}(M, \g)$ satisfying  
\begin{equation}
\label{swsw''}
r \left( \tbS_{L_1}(f  \Phi) \right)
= \bS_{L_1 + \delta L_1} (Z f \Phi + \bdel_{L_1}(f\Phi)),  
\end{equation}
where $f$ is now an arbitrary test density of compact support.
We can replace $f\Phi$ in the above formula by 
a sum $\sum \alpha_i f_i \Phi_i$ and 
differentiate the formula $n$ times with respect to 
to the parameters $\alpha_i$ (setting these parameters to zero afterwards). This gives us 
the desired identity~\eqref{VAV}. \qed

\section{The Renormalization Group in Curved Spacetime}

\subsection{Scaling of interacting fields}

As explained in the previous section, it is possible to give a
perturbative construction of the interacting quantum field theory that
defines the interacting field, its Wick products, and their time
ordered products as local, covariant fields.  The construction of this
theory depends on a prescription for defining Wick powers and their
time-ordered-products in the corresponding free theory. As also explained,
the definition of these quantities involves some ambiguities, and
consequently the definition of the interacting field theory is also
ambiguous. Nevertheless we showed in the previous subsection that
these ambiguities can be analyzed in much the same way as in the free
theory. The result of this analysis was summarized in eq.~\eqref{VAV}.

In the present section we want to investigate the behavior of the
interacting field, its Wick powers, and their time-ordered-products in
the interacting theory under a rescaling of the metric by a constant
conformal factor $\lambda$. As explained in the introduction, this
analysis corresponds to a definition of the renormalization group in
curved spacetime.

For the Wick powers and time-ordered-products in the free theory, the
scaling behavior was analyzed at the end of section 2 using the
``scaling map'', $\sigma_\lambda$, (introduced in eq.~\eqref{bdef}
above), which associates to every element of $\cW(M, \lambda^2 \g,
p(\lambda))$ a corresponding element of $\cW(M, \g, p)$,
where $p(\lambda) = (\lambda^{-2} m^2, \xi)$ are the rescaled coupling
constants. Choose an arbitrary, but fixed, prescription for defining
Wick powers and their time-ordered-products in the free theory 
that satisfy the axioms of \cite{hw1} and \cite{hw2}.
Let $\lambda$ be an arbitrary, but fixed, positive real number, and let
$\Phi$ be a Wick power with engineering dimension $d$. We define
\begin{equation}
\label{philambda}
^\lambda \Phi[M, \g, p](f) = \lambda^d \,\, \sigma_\lambda \left(
\Phi[M, \lambda^2 \g, p(\lambda)] (f) \right),
\end{equation}
and we similarly define $^\lambda T(\prod {^\lambda \Phi_i})[M, \g,
p]$.  It follows immediately that $^\lambda \Phi$ and $^\lambda
T(\prod {^\lambda \Phi_i})$ provide prescriptions for defining Wick
powers and their time-ordered-products that also satisfy all of the
axioms of \cite{hw1} and \cite{hw2}. As we have already noted, it then
follows that the relation of this new $\lambda$-dependent prescription
to the original prescription is of the form given by
eq.~\eqref{freesc} (but without the factors of $\lambda^{-d_T}$
occurring on the right side of that equation).

In order to analyze the scaling behavior of the fields in the
interacting theory defined by the interaction Lagrangian density
$\cL_1 = \kappa \varphi^4 \beps$, we proceed as follows.  Our new
$\lambda$-dependent prescription, eq.~(\ref{philambda}), for defining
Wick powers and their time-ordered-products for the free field gives
rise, via the construction of section 3.1, to a new
$\lambda$-dependent prescription for the perturbative construction of
the corresponding interacting fields, which we denote by $^\lambda
\Phi_{L_1}$ and $^\lambda T_{L_1}(\prod {^\lambda \Phi_i})$, respectively.
These quantities span an algebra of interacting fields denoted by
$^\lambda \cB_{L_1}(M, \g)$.  From the uniqueness result,
eq.~\eqref{VAV}, for the interacting Wick powers and their
time-ordered-products derived in the preceeding subsection we then
immediately get, for each $\lambda > 0$, a *-isomorphism
\begin{equation}
\j_\lambda: {^\lambda \cB}_{L_1}(M, \g) \to 
\cB_{L_1 + \delta L_1(\lambda)}(M, \g).
\end{equation}
Here,
$\delta\cL_1(\lambda)$ is the $\lambda$-dependent counterterm
Lagrangian of the form~\eqref{countert}, whose $\lambda$-dependent
coupling parameters are given by formal power series in $\kappa$.  The
coefficients in these power series are {\it polynomials} in $\ln
\lambda$ whose degree increases with $n$; for example
\begin{equation} 
\label{dm2}
\delta m^2(\lambda) = \sum_{n =1}^\infty p_n(\ln \lambda) \kappa^n, 
\end{equation}
where the $p_n$'s are polynomials\footnote{It is possible to derive
inequalities for the maximum degree of the polynomials $p_n$ as a
function of the order $n$ in perturbation theory.}.  

It is not difficult to see that the relation between $^\lambda
\Phi_{L_1}$ and $\Phi_{L_1}$ is simply
\begin{equation}
\label{philambdaL1}
^\lambda \Phi_{L_1}[M, \g, p](f) = \lambda^d \,\, \sigma_\lambda \left(
\Phi_{L_1}[M, \lambda^2 \g, p(\lambda)] (f) \right),
\end{equation}
where here we have again denoted by $\sigma_\lambda$ the obvious
extension of $\sigma_\lambda$ from $\cW$ to $\cX$. A similar formula
holds for the time-ordered-products of the interacting fields. Consequently,
if we compose
$\sigma_\lambda$ with $\j_\lambda$ constructed above, we obtain a
*-isomorphism $\ccR_\lambda = \j_\lambda \circ \sigma_\lambda$
\begin{equation}
\ccR_\lambda: \cB_{L_1}(M, \lambda^2 \g, p(\lambda)) \to \cB_{L_1 + 
\delta L_1(\lambda)}(M, \g, p), 
\label{Rgflow}
\end{equation}
where we indicate explicitly the dependence on the parameters $p$ in
the free theory. Since the scaling map $\sigma_\lambda$ in the free theory 
satisfies 
$\sigma_\lambda \circ \sigma_{\lambda'} = \sigma_{\lambda\lambda'}$, it follows that
\begin{equation}
\ccR_\lambda \circ \ccR_{\lambda'} = \ccR_{\lambda\lambda'}
\end{equation}

Using eq.~\eqref{VAV} we find that the 
action of $\ccR_\lambda$ on an interacting 
time-ordered-product in the algebra $\cB_{L_1}(M, \lambda^2 \g, p(\lambda))$ is 
given by 
\begin{multline}\label{tscale}
\ccR_\lambda \left( {T_{L_1}}\left(\prod_{i=1}^n   {\Phi_i} (f_i) \right) \right)
= \lambda^{-d_T} \,\, T_{L_1 + \delta L_1(\lambda)} \left( \prod_{i=1}^n
Z_i(\lambda) \Phi_i(f_i) \right) + \\ 
\lambda^{-d_T}
\sum_{P} T_{L_1 +
\delta L_1(\lambda)} \left(\prod_{I \in P} \bO_{|I|}(\lambda; \times_{i \in
I} f_i \Phi_i) \prod_{j \notin I \, \forall I \in P} Z_j(\lambda)
{\Phi_j}(f_j) \right).
\end{multline}
Here, the $\lambda$-dependent field strength renormalization factors,
$Z_i(\lambda)$, can be written as $Z_i(\lambda) = 1 +
\sum_{n\ge 1} z_{i,n}(\ln \lambda) \kappa^n$, where the coefficients
$z_{i,n}$ depend at most polynomially on $\ln \lambda$.  
The terms $\bO_n(\lambda; \times_i f_i \Phi_i)$ have
the same form as eq.~\eqref{VAV}, and each of the terms in the sum on
the right side of this equation is a formal power
series in $\kappa$, whose coefficients are geometric tensors times
polynomials in $\ln \lambda$. For the special case of the interacting field
$\varphi_{L_1}$, the above formula simplifies to
\begin{equation}
\ccR_\lambda(\varphi_{L_1}(f)) = \lambda^{-1} 
Z(\lambda) \varphi_{L_1 + \delta L_1(\lambda)}(f).
\end{equation}

Equation~\eqref{tscale} is our desired formula for the scaling
behavior of the fields in the interacting quantum field theory.
Although eq.~(\ref{tscale}) has many obvious similarities to the
corresponding formula eq.~\eqref{freesc} in the free theory, it should
be noted that there are a number of important differences, in parallel
with the differences in the general renormalization ambiguities of the
free and interacting fields (see section 3.2 above). Most prominently,
in the free field theory, the scaling relations~\eqref{freesc} relate
rescaled time-ordered-products to the unscaled time-ordered-products
defined with respect to the ``{\it same} Lagrangian'', whereas the
scaling relations eq.~\eqref{tscale} in the interacting
theory\footnote{For the classical interacting field, the scaling
relations also do not involve a modification of the interaction
Lagrangian, as can be seen from the fact that the classical Lagrangian
$\cL$ (and the corresponding classical nonlinear equations of motion)
is manifestly invariant under transformation $\g \to \lambda^2 \g$,
$\varphi \to \lambda^{-1} \varphi$, $m^2 \to \lambda^{-2} m^2$ and
$\xi \to \xi$. This can also be seen, more indirectly, in present
formalism if one keeps explicitly the dependence of our constructions
on $\hbar$, so that the corresponding classical theory corresponds to
the limit $\hbar \to 0$. This is most naturally done by introducing
$\hbar$ as an explicit parameter in our definition of the product
``$\star$'', eq.~\eqref{omegaprod}, in our algebra $\cW$ (and likewise
$\mathcal X$), see~\cite{fd}.} relate the rescaled
time-ordered-products for the interaction Lagrangian $\cL_1$, to the
unscaled time-ordered-products defined with respect to the interaction
Lagrangian $\cL_1(\lambda) = \cL_1 + \delta \cL_1(\lambda)$. Another
important difference between the scaling relations~\eqref{tscale} and
~\eqref{freesc} is the occurrence of the field strength
renormalization factors, $Z_i(\lambda)$, in the interacting field
theory, while such factors are absent in the free theory. As a
consequence, the interacting fields do not in general have an almost
homogeneous scaling behavior.

Given any fixed renormalization prescription, eq.~(\ref{Rgflow}) shows
that the theory defined for the rescaled metric and rescaled
parameters of the free theory is equivalent to the original theory
with a modified Lagrangian $\delta \cL_1$.
The $\lambda$-dependence of the parameters $\delta m^2 (\lambda)$, $\delta
z(\lambda)$, $\delta \xi (\lambda)$, and $\delta \kappa (\lambda)$ in
$\delta \cL_1(\lambda)$ define the {\em renormalization group flow} 
of the theory. As already mentioned $\delta m^2$, $\delta
z$, $\delta \xi $, and $\delta \kappa$ are formal power series in 
$\kappa$. These quantities also depend upon the parameters appearing in 
$\cL_0$, so $\delta m^2$, $\delta z$, $\delta \xi $, and $\delta \kappa$ 
should be viewed as effectively being functions of $\kappa, m^2$, and $\xi$, as well as 
of $\lambda$. However, it should be noted that the renormalization group
flow is independent of the spacetime metric $\g$.

The physical meaning of the renormalization group flow can perhaps be
best explained by imagining that a quantum field theory textbook from
an ancient civilization has been discovered.  This textbook contains a
complete description of perturbative renormalization theory for the
scalar field (\ref{L}) as well as complete instructions on how to
build apparatuses to prepare states of the theory and to make
measurements (see the discussion at the end of section 3.1). It also
records the results of these measurements and compares them with
theoretical predictions (to some appropriately high order in
perturbation theory), thereby fixing the parameters of the
theory. However, the one piece of information that is missing is the
system of units used by the ancient civilization; in other words, the
lengthscale, $l$, used by the ancient civilization to define the
fundamental unit of length (in terms of which other units, such as
mass, are defined in the standard way) is not presently known. This
lengthscale enters both the renormalization prescription given in the
book (since, the specification of a particular locally constructed
Hadamard parametrix and the renormalization prescription for defining
time-ordered-products require a specification of a unit of length) as
well as the instructions for building the apparatuses and making the
measurements. Suppose, now, that a physicist from the present era
tries to verify the experimental claims made in the book. He makes a
guess, $l'$, as to the value of $l$, which, as it turns out, differs
from $l$ by a factor of $\lambda^{-1}$, i.e., $l' = l/\lambda$. Since
the present-day physicist will normalize the spacetime metric so that
a rod of length $l'$ will have unit length (whereas the ancient
civilization assumed that a rod of length $l$ has unit length), the
spacetime metric $\g'$ used by the present-day physicist will differ
from the metric $\g$ that would have been used by the ancient
civilization by $\g' = \lambda^2 \g$. Correspondingly, all of the
experimental apparatuses built by the present-day physicist will be a
factor of $\lambda$ smaller in all linear dimensions than intended by
the author of the ancient textbook. When the present day physicist
completes his experiments, he will find that his results disagree with
the results reported in the book. He will find that this disagreement
will be alleviated he compares his results to the theoretical
predictions obtained from the renormalization prescription given in
the book by using the mass parameter $m^{\prime } = \lambda^{-1}m$ in $\cL_0$
rather than $m$, but disagreements will still remain. However, if, in
addition to the substitution $m^{\prime } = \lambda^{-1} m$ in $\cL_0$, the
present-day physicist also modifies the interaction Lagrangian $\cL_1$
by eq.~(\ref{countert}) (with $\delta z, \delta m^2$, etc. given by eq.~(\ref{dm2})), 
then he will find exact agreement with the theoretical
predictions obtained from the renormalization prescription given in
the book, provided that he also redefines the field variables in accordance
with the *-isomorphism $\ccR_\lambda$ given by eq.~(\ref{tscale}).
In other words, when the properties of the scalar field are
investigated on a scale different from that used by the ancient
civilization, its properties will be found to differ by a ``running of
coupling constants'' in the interaction Lagrangian.

The quantity $\delta \kappa (\lambda)$ can be viewed as modifying the
nonlinear coupling parameter $\kappa$ appearing in the original
interaction Lagrangian $\cL_1$. However, it should be noted that the
quantities $\delta m^2 (\lambda)$, $\delta z(\lambda)$, and $\delta
\xi (\lambda)$ all correspond to parameters appearing in the original
free Lagrangian, $\cL_0$, rather than $\cL_1$. It would be natural to
try to interpret these terms in $\cL_1$ as corresponding to changes in
the coupling constants $m^2$, $z = 1$, and $\xi$ appearing in
$\cL_0$. However, we do not know how to justify 
such an interpretation because we have only constructed the
interacting theory at the level of a formal perturbation
expansion. Therefore, we cannot compare an interacting theory based on
the free Lagrangian $\cL_0$ with an interacting theory based on the
free Lagrangian $\cL_0 + \delta \cL_0$, where 
$\delta \cL_0 = \frac{1}{2}[\delta z (\nabla\varphi)^2 + 
\delta \xi R \varphi^2 + \delta m^2 \varphi^2]\beps$.

Finally, as we have already noted, the renormalization group flow occurs in the
parameter space of the theory and is independent of the spacetime
metric. Thus, in order to calculate (or measure) the renormalization
group flow, it suffices to restrict attention to a single spacetime,
provided that the spacetime is not so special that possible curvature
couplings do not occur. Thus, for example, in the theory with
Lagrangian~\eqref{L}, the only coupling to curvature occurs in the
term $\xi R \varphi^2$, so it would suffice to calculate the
renormalization group flow in any spacetime with nonvanishing scalar
curvature. We will indicate how to calculate renormalization group
flow in curved spacetime in terms of Feynman diagrams in appendix~B.
However, we point out here that a great deal of information
about the renormalization group flow can be deduced from dimensional
considerations as well as from some simple properties that hold in
special spacetimes\footnote{We are indebted to K.-H.~Rehren, C.J.~ Fewster,
and K.~Fredenhagen for bringing this point to our attention.}. From
dimensional considerations alone, it follows that the dependence of
$\delta m^2 (\lambda)$, $\delta z(\lambda)$, $\delta \xi (\lambda),
\delta \kappa(\lambda)$ on the parameters $m^2$, $\xi$, and
$\kappa$ must be of the form $\delta m^2(\lambda) = m^2 F_{m^2}(\lambda, \xi,
\kappa)$, $\delta z(\lambda) = F_z(\lambda, \xi, \kappa)$, $\delta \xi(\lambda) =
F_\xi(\lambda, \xi, \kappa), \delta \kappa(\lambda) = F_\kappa(\lambda,
\xi, \kappa)$. However, it is possible (and would be very natural) to
choose a prescription for defining free field Wick products and their
time-ordered-products in an arbitrary spacetime such that in the
special case of Minkowski spacetime, this prescription does not depend
upon the irrelvant parameter $\xi$. It follows immediately that with 
such a renormalization prescription, the
renormalization group flow cannot depend upon $\xi$ in Minkowski
spacetime and, therefore---since the flow is independent of the
spacetime metric---the flow cannot depend upon $\xi$ in any
spacetime. More generally, it is possible (and would be very natural)
to choose a prescription for defining free field Wick products and
their time-ordered-products in an arbitrary spacetime such that in the
special case of a spacetime with constant scalar curvature $R$ (such as deSitter spacetime), 
the only dependence of the
prescription on the parameters $m^2$ and $\xi$ occurs in the
combination $m^2 + \xi R$. This condition implies that (in all
spacetimes), the renomalization group flow must take the form
\begin{eqnarray}
\delta m^2 &=& m^2 G_1(\lambda,\kappa) \nonumber \\
\delta \xi &=& \xi G_1(\lambda,\kappa) + G_2(\lambda,\kappa) \nonumber \\
\delta z   &=& G_3(\lambda,\kappa) \nonumber \\
\delta \kappa &=& G_4(\lambda,\kappa)
\label{RGflow} 
\end{eqnarray}
The functions $G_1$, $G_3$, and $G_4$ can all be determined by calculations
done entirely in Minkowski
spacetime; the function $G_2$ cannot be determined by calculations in
Minkowski spacetime but could be determined by calculations done, e.g., 
in deSitter spacetime.

\subsection{Fixed points, essential vs. inessential coupling parameters}

In the previous section we have seen that a rescaling of the spacetime
metric by a constant conformal factor, $\g \to \lambda^2 \g$, (a
``change of length scale'') gives rise to different definitions of the
interacting field theory. The relation between the definitions of
the field theory at different length scales is given by the
renormalization group. It is of interest to ask at what points in the
parameter space of the theory the definition of a field theory is
actually ``independent'' of the scale at which it is defined. Such
points are usually referred to as ``fixed points''.

Naively, one might attempt to define a fixed point as a point in
parameter space at which the $\lambda$-derivatives of $\delta m^2
(\lambda)$, $\delta z(\lambda)$, $\delta \xi (\lambda)$, and $\delta
\kappa(\lambda)$ all vanish. However, this definition would be too
restrictive because it excludes points where the renormalization group
flow is nonvanishing but corresponds merely to a redefinition of field
variables. One would like to define the notion of fixed points so that
it also includes points in parameter space where the renormalization group
flow is nonvanishing but is tangent to a trivial flow corresponding to
a field redefinition.

To see more explicitly the nature of such trivial flows, consider a
field theory with Lagrangian $\cL(\varphi)$ and consider a mapping
$\varphi \to F(\varphi)$ on field space such that $F(\varphi)(x)$ 
depends only on $\varphi(x)$ and finitely many of its covariant derivatives
at the point $x$. Then, although the Lagrangian $\cL(\varphi)$ and
$\cL(F(\varphi))$ may look very different (i.e., different kinds of
couplings and different values of coupling parameters), they
nevertheless would define an equivalent classical field theory. Thus,
at the classical level, there is a wide class of trivial flows in
parameter space that correspond to field redefinitions. However, the
situation is far more restrictive for a field with Lagrangian
(\ref{L}) if we want the field redefinition to keep the Lagrangian in
a perturbatively renormalizable form. It is not difficult to see that
(in 4 dimensions) this leaves us only with the possibility to multiply
the field by a constant, i.e., the only possible form of $F$ is
$F(\varphi) = s \varphi$. The new classical
Lagrangian $\cL(s) \equiv \cL(F(\varphi))$ is then
\begin{equation}
\cL(s) =  \frac{1}{2}[s^2 (\nabla \varphi)^2 + s^2(m^2 + \xi R) \varphi^2 + 
s^4 \kappa \varphi^4 ]\beps. 
\end{equation}
If one splits this Lagrangian into its free and interacting parts via
$\cL(s)= \cL_0 + \cL_1(s)$ with $\cL_0 = \frac{1}{2}[(\nabla
\varphi)^2 + m^2 \varphi^2 + \xi R \varphi^2]\beps$, the interaction
Lagrangian takes the form
\begin{equation}
\label{l1s}
\cL_1(s) = 
\frac{1}{2}[(s^2 - 1)(\nabla \varphi)^2 + (s^2 - 1)(m^2 + \xi R) \varphi^2 + 
s^4 \kappa \varphi^4] \beps.  
\end{equation}
Therefore, one might expect that the ``one-parameter flow'' defined by
eq.~(\ref{l1s})---with $s$ taken to be an arbitrary power series in
$\kappa$---would correspond to a trivial flow in the parameter space
of the theory in the sense that the theory constructed from the
interaction Lagrangian $\cL_1(s)$ would be equivalent to the theory
constructed from the original interaction Lagrangian $\cL_1 =
\frac{1}{2} \kappa \varphi^4 \beps$.

However, the actual situation is somewhat more complicated than the above
considerations might suggest. The theories constructed from the
interaction Lagrangians $\cL_1(s)$ and $\cL_1$ will depend upon the
specific choice of renormalization prescription, and, for any given
prescription, we see no reason why these two theories need be
equivalent. Indeed, it appears far from clear that there exists any
renormalization prescription that gives equivalence of the two
theories. Nevertheless, we shall now show that, for any fixed
renormalization prescription, there exists {\em some} one-parameter
family of interaction Lagrangians, $\cK_1(s)$, such that the theories
constructed from $\cK_1(s)$ are equivalent to the theory constructed
from $\cL_1$ in the sense that the algebras $\cB_{K_1(s)}(M,
\g)$ and $\cB_{L_1}(M, \g)$ are isomorphic. Furthermore, the action of
this isomorphism on the interacting field corresponds to the simple
field redefinition $F(\varphi) = N(s)\varphi$, where $N(s)$ is 
a formal power series with the propery $N(s=1) = 1$. The precise
statement of this result is as follows:

\begin{thm}
Let $s = 1 + \sum_{i \ge 1} s_i \kappa^i$ be a formal power series in $\kappa$ with real
coefficients. Then there exists an interaction Lagrangian $\cK_1(s)$
of the same form as the original Lagrangian, a formal power series
$N(s)$ and a *-isomorphism $\rho_s: \cB_{L_1}(M, \g) \to \cB_{K_1(s)}(M,
\g)$ such that
\begin{equation}
\label{1pe''}
\rho_s \left[ \varphi_{L_1}(f) \right] = N(s)
\varphi_{K_1(s)} (f)
\end{equation}
for all $f \in \cD(M)$, and such that $N(s = 1) = 1$ and $\cK_1(s=1) = \cL_1$. 
\end{thm}

A proof of this theorem is given in appendix~A.

According to the above theorem, it is natural to view the interaction
Lagrangians $\cL_1$ and $\cK_1(s)$ as defining the same quantum field
theories and $\rho_s$ as implementing the field redefinition. If we
choose coordinates on the space of parameters in the Lagrangian so
that the coordinate vector field of one of the coordinates is tangent to the flow defined by
$\cK_1(s)$, then we refer to this coordinate as an {\em inessential
parameter} of the theory (see, e.g., \cite{w}). We define a fixed
point of the renormalization group flow to be a point at which only
the inessential parameter changes under the flow. More precisely, if
$\lambda \to \cL_1(\lambda)$ is the renormalization group flow, then
we say that we are at a fixed point if there is a 1-parameter family
$\lambda \to s(\lambda)$ such that
\begin{equation}
\cL_1(\lambda) = \cK_1(s(\lambda)) \quad \text{for all 
$\lambda > 0$.} 
\end{equation}
This relation can be differentiated with respect to $\ln \lambda$,
thereby relating a fixed point to a zero of a suitably defined
$\beta$-function.
For this, we write $\cL_1(\lambda) = \cL_1 + \delta \cL_1(\lambda)$, and 
$\cK_1(s) = \cL_1 + \delta \cK_1(s)$, and we denote the parameters
in $\delta \cL_1(\lambda)$ by $\delta z(\lambda), \delta \kappa(\lambda)$ etc.
and the parameters in $\delta \cK_1(s)$ by 
$\delta \tilde z(s), \delta \tilde \kappa(s)$ etc. We 
define\footnote{If $\cK_1(s)$ were actually of the form
(\ref{l1s}), then the $\beta$-function for $\kappa$ would be given by
$\beta_\kappa \equiv \frac{\partial}{\partial \ln \lambda} (\delta
\kappa(\lambda) - 2 \kappa \delta z(\lambda))|_{\lambda = 1}$} 
\begin{equation}
\label{bfunction}
\beta_\kappa \equiv \frac{\partial}{\partial \ln \lambda} \delta \kappa
- \frac{\partial}{\partial s} \delta \tilde \kappa \left( 
\frac{\partial}{\partial s} \delta \tilde z \right)^{-1} 
\frac{\partial}{\partial \ln \lambda} \delta z \Bigg|_{\lambda = s = 1}.
\end{equation}
Then a fixed point\footnote{It should be noted that the interacting theory has
been constructed only at the level of a formal perturbation expansion, it 
will not be possible to reliably determine fixed points unless they occur near 
$\kappa = 0$.} 
corresponds to a zero of $\beta_\kappa$ (together with a zero of similarly defined 
beta functions $\beta_{m^2}, \beta_\xi$).

{\bf Acknowlegements:} This work was supported in part by NSF grant PHY00-90138 to the
University of Chicago. Part of this research was carried out during the
program on Quantum Field Theory in Curved Spacetime at the Erwin
Schr\"odinger Institute, and we wish to thank the Erwin Schr\"odinger
Institute for its hospitality.


\appendix
\section{Appendix A}

In this appendix we give a proof of theorem~4.1.
Mainly for notational simplicity, we will assume throughout this proof that $\xi = m^2 = 0$, so that 
$\cL_0 = \frac{1}{2}(\nabla \varphi)^2 \beps$; the general case can be treated in exactly the same way. 
Consider the Lagrangian density $\delta \cL_0 = \frac{1}{2}\delta s(\nabla \varphi)^2 \beps$ with $\delta s = 
s^2 - 1$, and 
a cutoff function $\theta$ which is equal to 1 in a neighborhood of the closure $\bar V$ of 
a globally hyperbolic neighborhood $V$ with compact closure and with a Cauchy surface of the form $\Sigma \cap V$, 
where $\Sigma$ is a Cauchy surface for $M$. Although $\delta \cL_0$ is, of course, only quadratic in the 
field $\varphi$, we may consider it as an ``interaction Lagrangian,''  
and we can define, by eqs.~\eqref{intfield} respectively \eqref{inttop} (with $\cL_1$ in those
equations replaced by $\delta \cL_0$), the corresponding ``interacting'' fields  
as formal power series in $\delta s$ (or, more properly, as formal power series in $\kappa$, 
since $s$ itself is a formal power series in $\kappa$). 

The first step in our proof is to show that the ``interacting
fields'' $\varphi_{\theta \delta L_0}(f)$ with $f$ a smooth test density of compact support in $V$
satisfy exactly the same algebraic relations as the fields $s^{-1}\varphi(f)$. Furthermore, 
we show that the ``interacting time-ordered-products'' 
$T_{\theta \delta L_0}(\prod \Phi_i(f_i))$ (with the support of $f_i$ contained in 
$V$) satisfy commutation relations with the field $\varphi_{\theta \delta L_0}(f)$ that have exactly the 
same form as the commutation relations of $s^{-N} T(\prod \Phi_i(f_i))$ with $s^{-1}\varphi(f)$ given in~\cite{hw2}, 
where $N$ is the number of free field factors in the time-ordered-product. We formulate this result as a lemma.

\begin{lemma} 
For all smooth test densities with support in $V$, we have that 
\begin{equation}
\label{kg}
\varphi_{\theta \delta L_0}(\square f) = 0, \quad 
\varphi_{\theta \delta L_0}(f)^* = \varphi_{\theta \delta L_0}(\bar f),  
\quad [\varphi_{\theta \delta L_0}(f_1), \varphi_{\theta \delta L_0}(f_2)] = is^{-2}\Delta(f_1, f_2) \myid 
\end{equation}
in the sense of formal power series\footnote{For example, $s^{-1}$ is defined as the formal 
power series $\sum_n (-1)^n (\sum_{i \ge 1} s_i \kappa^i)^n$.} in $\kappa$. More generally it holds that
\begin{multline}
\label{com}
\left[ T_{\theta \delta L_0}(\prod_{i=1}^n \Phi_i(f_i)), \varphi_{\theta \delta L_0}(f_{n+1}) \right] = \\
s^{-2} \sum_{j = 1}^n 
T_{\theta \delta L_0}(\Phi_1(f_1) 
\dots i\sum_{(a)}\frac{\partial \Phi_j}{\partial \nabla_{(a)}\varphi}(f_{n+1} \Delta_{(a)} f_j) 
\dots \Phi_n(f_n)),    
\end{multline}
where $f_{n+1} \Delta_{(a)} f_j$ was defined in
eq.~\eqref{hdf1}.
\end{lemma}
\begin{proof}
In order to prove the first relation in eq.~\eqref{kg}, we first expand 
\begin{equation}
\varphi_{\theta \delta L_0}(f) 
= \varphi(f) + \sum_{n\ge 1} \frac{(i\delta s)^n}{n!} R(f \varphi; 
\underbrace{\theta\cL_0, \dots, \theta \cL_0}_{n \,\, factors}).    
\label{varphil0}
\end{equation}
Since $\cL_0$ is only quadratic in the field $\varphi$, the totally
retarded products~\eqref{varphil0} can be given in closed form in 
terms of the retarded Green's function $\Delta_{ret}$ for $\square$, 
\begin{multline}
\label{string1}
R(\varphi(x); \prod_{i = 1}^n \cL_0(y_i)) = i^n
\sum_{i_1 \cdots i_n} \Delta_{ret}(x, y_{i_1}) \lnabla \rnabla \times \\
\Delta_{ret}(y_{i_1}, y_{i_2}) \lnabla \rnabla \cdots \Delta_{ret}(y_{i_{n-1})}, y_{i_n}) 
\lnabla \rnabla \varphi(y_{i_n}),    
\end{multline}
where the summation over the spacetime index has been suppressed in the expression $\lnabla\rnabla$.
We now use this 
expression to analyze the oparator $R(\square f \varphi; \times^n \theta \delta  \cL_0)$, where 
$f$ is a test density supported in $V$. In order to do this, we perform the following steps: 
We use $\square \Delta_{ret} = \delta$ to turn the first retarded Green's function on the right side of
eq.~\eqref{string1} into a delta-function. We then use that $\theta$ is 1 in $V$ and that $f$ has support in $V$
and perform $n$ successive partial integrations in order to turn the $\lnabla \rnabla$ derivatives
into $\rnabla \rnabla$ derivatives which will now hit a single retarded Green's function, thus resulting
each time in a new delta-function. If this is done, then one obtains 
$R(\square f \varphi; \times^n \theta \delta \cL_0) = 0$, 
thereby proving the first equation in \eqref{kg}. The second equation in~\eqref{kg}
follows from the unitarity of the relative $S$-matrix $\bS_{\theta \delta L_0}(f\varphi)$ for real-valued 
$f$. 

We will demonstrate eq.~\eqref{com} in the case of Wick powers of the form $\varphi^k$; Wick powers with 
derivatives and time-ordered-products can be treated similarly. 
The proof of the last relation in eq.~\eqref{kg} is included as the special case $k = 1$. 
Our starting point is the relation~\cite{fd}\footnote{A general formula of this kind which holds 
within the LSZ-framework in Minkowski spacetime was first given by~\cite{glz}.}
\begin{multline}
\label{commm}
\left[\varphi^k_{\theta \delta L_0}(x_1), \varphi^{}_{\theta \delta L_0}(x_2)\right]
= 
\sum_{n \ge 0} \frac{(i\delta s)^n}{n!} \int_{M^{\times n}} \prod_j 
\theta(y_j) \times \\
\left( R(\varphi^k(x_1); \varphi(x_2) \prod_{j=1}^n \cL_0(y_j)) 
- 
R(\varphi(x_2); \varphi^k(x_1) \prod_{j=1}^n \cL_0(y_j)) \right),  
\end{multline}
where the integral is over the ``$y$''-variables.
We will now simplify the terms under the sum in the above equation, starting with the 
terms $R(\varphi^k(x_1); \varphi(x_2) \prod_{j=1}^n \cL_0(y_j))$. For this, we use the fact 
that the time ordered products with a factor $\varphi$ can be shown to satisfy the following 
requirement in addition to any other requirements imposed so far\footnote{A proof of this 
equation for Minkowski spacetime appears in~\cite{fd1}. This proof can be generalized to 
curved spacetimes by suitably modifying the constructions of time ordered products given in~\cite{hw2}.}: 
\begin{equation}
(\square)_x T(\varphi(x) \prod_{j=1}^n \Phi_j(y_j))
= i\sum_{j=1}^n \sum_{(b)} \nabla_{(b)} \delta(y_j,x) T(\Phi_1(y_1) \cdots
\frac{\partial\Phi_j}{\partial \nabla_{(b)}\varphi}(y_j) \cdots \Phi_n(y_n))
\end{equation}
for all fields $\Phi_j$. It can be seen that this implies 
\begin{multline}
\label{N1}
R(\varphi^k(x_1) ; \varphi(x_2) \prod_{j=1}^n \cL_0(y_j)) 
= i 
\sum_{l=1}^n  \nabla_a \Delta_{ret}(y_l, x_2) R(\varphi^k(x_1) ; \nabla^a \varphi 
(y_l) \prod_{j \neq l}  \cL_0(y_j)) \\
+ i\Delta_{ret}(x_1, x_2) R(\frac{\partial \varphi^{k}}{\partial \varphi}(x_2); 
\prod_{j=1}^n  \cL_0(y_j)) 
\end{multline}
for the retarded products appearing in eq.~\eqref{commm}.
Now the retarded products in the sum on the right side of eq.~\eqref{N1} again 
contain a factor $\varphi$, and we can a similar argument as above to 
further simplyfy each of these terms. Repeating this procedure $n$ times, we can rewrite
the right side of eq.~\eqref{N1} as 
\begin{multline}
\label{string}
= i \sum_{N = 0}^n i^N \sum_{l_1 \cdots l_N}
\Delta_{ret}(x_1, y_{l_1}) \lnabla \rnabla 
\Delta_{ret}(y_{l_1}, y_{l_2}) \lnabla \rnabla \cdots \Delta_{ret}(y_{l_N}, x_2) \\
\times R(\frac{\partial \varphi^k}{\partial \varphi}(x_2); \prod_{j \neq l_1, \dots, l_N} \cL_0(y_j)).   
\end{multline}
The second term  
$R(\varphi(x_2); \varphi^k(x_1) \prod_{j=1}^n \cL_0(y_j))$ under the sum in eq.~\eqref{commm} can 
be written in the form of expression~\eqref{string} with $x_1$ and $x_2$ exchanged. 
We now substitute these expressions back into 
\eqref{commm} and perform the following steps: We use that $x_1, x_2 \in V$, that 
$\theta \equiv 1$ on $V$ and the support property $\supp \Delta_{ret} \subset \{(x_1, x_2) \in M \times M \mid
x_1 \in J^+(x_2) \}$ to bring each of turn each of the $\lnabla \rnabla$ derivatives on the variables 
$y_{l_j}$ into a $\rnabla \rnabla$ derivative acting on a single retarded Green's function via a partial integration.
We then use that $\square \Delta_{ret} = \delta$ and use these delta-functions to get rid of the 
string of retarded Green's functions in~\eqref{string}. We now exploit the relation
$\Delta_{ret}(x_1, x_2) = \Delta_{adv}(x_2, x_1)$ (with $\Delta_{adv}$ the advanced Green's function), 
as well as $\Delta = \Delta_{adv}-\Delta_{ret}$, which enables one to get rid of all 
retarded Green's functions in favor of commutator functions. We finally collect similar terms and 
use the geometric series $\sum_{N=0}^{\infty} (\delta s)^N = s^{-2}$ (here it must be used that 
$s$ has the special form $1 + \sum_{i \ge 1} s_i \kappa^i$, or else the formal power series 
$s^{-2}$ is not well-defined). If all this 
is done, then one obtains \eqref{com} for the special case of a Wick product of the form $\varphi^k$.
\end{proof}

It follows from eqs.~\eqref{kg} and~\eqref{com} that the linear map  
\begin{equation}
\label{iodefa}
\rho_{\theta} \big[\varphi(f_1) \star \cdots \star \varphi(f_n) \big] \equiv s^n 
\, \varphi_{\theta \delta L_0}(f_1) \star \cdots \star \varphi_{\theta \delta L_0}(f_n) 
\end{equation}
defines a *-homomorphism from the canonical commutation relation 
algebra $\cA(V, \g)$ into the subalgebra of 
${\mathcal X}(M, \g)$ spanned by products of the fields $\varphi_{\theta \delta L_0}(f)$, 
where $f$ is an arbitrary test density supported in $V$. Since the algebra $\cA(V, \g)$ is
simple, $\rho_\theta$ is injective. It is possible to see that 
the homomorphism $\rho_\theta$ can be extended by continuity\footnote{
It was shown in~\cite{hw1} that the H\"ormander topology on the spaces ${\mathcal E}'_{\sym}(M^{\times n})$
(see eq.~\eqref{esymdef}) induces a natural topology on the algebra $\cW(V, \g)$ and likewise on the algebra $\cX(M, \g)$. 
It can then be seen that the map $\rho_\theta$ defined in eq.~\eqref{iodefa} is continuous with respect to 
this topology.} to a unique *-homorphism from $\cW(V, \g)$, (and therefore also from $\cX(V, \g)$) to 
$\cX(M, \g)$. We will denote this extension by the same symbol $\rho_\theta$. 

We will now construct for any set of test densities $f_i$ of compact support in $V$
and for any set of fields $\Phi_i \in \cV$ an element $F(s; \times_i f_i \Phi_i) \in
{\mathcal X}(V, \g)$ such that 
\begin{equation}
\label{Ttilde}
\rho_{\theta} \left[ F(s; \times_{i=1}^n f_i \Phi_i) \right] = s^{N} \,
T_{\theta \delta L_0} (\prod_{i = 1}^n \Phi_i(f_i)),    
\end{equation}
where $N$ is the number of factors of $\varphi$ in the time-ordered-product. Furthermore, we claim that 
quantities $F(s; \times_i f_i \Phi_i)$ are independent of the particular choice of $\theta$ and $V$
and define in fact a new, $s$-dependent prescription for 
defining time-ordered-products in the free theory, i.e. that 
\begin{equation}
\tilde T(\prod_{i=1}^n \tilde \Phi_i(f_i)) \equiv F(s; \times_{i=1}^n f_i \Phi_i)
\end{equation}
satisfies all the requirements of our uniqueness theorem for time-ordered-products in the 
free theory. 

Before we sketch the proof of eq.~\eqref{Ttilde} and the claims following that equation, we would like
to mention that we see no reason obvious why the prescription $\tilde T$ should coincide with the 
original prescription $T$. As we will see below, the possible failure of $\tilde T$ to coincide with $T$
is the reason why the Lagrangian $\cK_1(s)$ in the theorem need not have the simple form 
expected from the classical theory. 

It follows from the relation 
\begin{equation}
\rho_{\theta'} = {\rm Ad}(U(\theta',\theta)) \circ \rho_\theta
\end{equation}
(with $U(\theta,\theta')$ defined as in eq.~\eqref{Udef}, but with $\cL_1$ in that equation 
replaced by $\delta \cL_0$) that if elements $F(s; \times_i f_i \Phi_i)$
satisfying eq.~\eqref{Ttilde} exist, then they must be independent of $\theta$. We now 
explain how to construct these elements.
By definition of $\rho_\theta$ given in eq.~\eqref{iodefa} we already know that 
eq.~\eqref{Ttilde} holds for the field $s\varphi_{\theta \delta L_0}(f)$ with $F(s; f\varphi)$
given by $\varphi(f)$ in that case.
The construction of $F(s; \times_i f_i \Phi_i)$ for a general time-ordered-product $s^N 
T_{\theta \delta L_0}(\prod \Phi_i(f_i))$ is as follows: 
On the algebra $\cW(M, \g)$, we consider, for all $t_i \in {\mathcal E}'(M, \g)$, 
the (commutative, associative) product\footnote{
If the $t_i$ are given by smooth densities $f_i$ on $M$, then the product
$W([f_1], \dots, [f_n])$ corresponds to the normal ordered product 
$\lno \varphi(f_1) \cdots \varphi(f_n) \rno_\omega$, where the normal ordering is done with respect to 
the quasifree state $\omega$ used in the definition of the algebra $\cW$.}   
\begin{equation}
\mbox{\huge $\times$}^n \cW(M,\g)
\to \cW(M, \g), \quad
\times_{i = 1}^n [t_i] \to 
W(\times_{i = 1}^n [t_i]) \equiv [t_1 \otimes_{\sym} \cdots \otimes_{\sym} t_n].
\end{equation} 
We also denote by $W$ the corresponding product on $\mathcal X(M, \g)$ when each $t_i$ is a formal 
power series in $\kappa$ with coefficients in ${\mathcal E}'(M, \g)$. 
Then it follows from the third equation in~\eqref{kg} that, within $V$, we have
\begin{equation}\label{wcom}
[W( \times_{k=1}^n \varphi_{\theta \delta L_0}(x_k)), \varphi_{\theta \delta L_0}(x_{n+1})]
= s^{-2} \sum_{k=1}^n i\Delta(x_k, x_{n+1}) W(\times_{j \neq k} \varphi_{\theta \delta L_0}(x_j)).
\end{equation} 
Since the time-ordered-products $T_{\theta \delta L_0}(\prod \Phi_i(f_i))$ satisfy similar commutation 
relations with the field $\varphi_{\theta \delta L_0}(f)$ (see eq.~\eqref{com}), it is possible to prove that, within $V$, these
time-ordered-products can expanded in terms of the products $W(\times_i \varphi_{\theta \delta L_0}(x_i))$
in a manner analogous to the usual Wick expansion,  
\begin{multline}
\label{wexpa}
T_{\theta \delta L_0}(\prod_{i=1}^n \varphi^{k_i}(x_i)) = 
\sum_{j \le k} {k \choose j}
\tau_{k_1-j_1 \dots k_n-j_n}(x_1, \dots, x_n) \times \\
W(\underbrace{\varphi_{\theta \delta L_0}(x_1), \dots, \varphi_{\theta \delta L_0}(x_1)}_{j_1 \,\, times}, \dots,
\underbrace{\varphi_{\theta \delta L_0}(x_n), \dots, \varphi_{\theta \delta L_0}(x_n)}_{j_n \,\, times}),  
\end{multline}
where the coefficients $\tau_{k_1-j_1 \dots k_n-j_n}$ are 
distributional and  
we use a multi-index notation $j = (j_1, \dots, j_n)$, $j! = \prod j_i!$, etc.
The proof of this statement is similar to 
the proof of the Wick expansion for the time-ordered-products in the free field theory given 
in~\cite{hw2}. Namely, we assume inductively that eq.~\eqref{wexpa} has been demonstrated for 
all multi indices $k$ with $|k| = \sum k_i < m$. In  order to prove it for a multi index
$k$ with $|k| = m$, we consider the expression
\begin{multline}
\label{wexpa'}
D_\theta(x_1, \dots, x_n) = T_{\theta \delta L_0}(\prod_{i=1}^n \varphi^{k_i}(x_i)) -
\sum_{0 \neq j \le k} {k \choose j}
\tau_{k_1-j_1 \dots k_n-j_n}(x_1, \dots, x_n) \times \\
W(\underbrace{\varphi_{\theta \delta L_0}(x_1), \dots, \varphi_{\theta \delta L_0}(x_1)}_{j_1 \,\, times}, \dots,
\underbrace{\varphi_{\theta \delta L_0}(x_n), \dots, \varphi_{\theta \delta L_0}(x_n)}_{j_n \,\, times}),  
\end{multline}
where the only term $\tau_{k_1 \dots k_n}$ that is not yet known by the induction hypothesis has 
been omitted from the sum in~\eqref{wexpa'}. The commutation relations for the individual terms on the 
right side of this equation now imply the commutation relation 
$[D_\theta(x_1, \dots, x_n), \varphi_{\theta \delta L_0}(y)] = 0$
within $V$. The above statements will still be true for a suitable $V$ containing a neighborhood of some
Cauchy surface $\Sigma$ of $M$. In this case, one can easily prove using eq.~\eqref{varphil0} and 
the above commutation relation that $D_{\theta}$ must in fact be a multiple of the identity. 
We define $\tau_{k_1 \dots k_n}$ to be this multiple. 
 
The products on the right side of eq.~\eqref{wexpa} 
can be written in terms of ordinary products using the formula
\begin{equation}
\label{wexpb}
W(\times_{i = 1}^N \varphi_{\theta \delta L_0}(x_i)) = 
\sum_{P} \prod_{j \notin I \, \forall I \in P}
\varphi_{\theta \delta L_0}(x_{j}) \prod_{P \owns I = 
\{i_1, i_2\}} \omega_{\theta \delta L_0}(x_{i_1},x_{i_2}).
\end{equation}
where $P$ runs over all sets of mutually disjoint subsets 
$I = \{i_1, i_2\}$ of  $\{1, \dots, N\}$ with 2 elements and where 
$\omega_{\theta \delta L_0}(x_1, x_2) = \omega(\varphi_{\theta \delta L_0}(x_1)
\varphi_{\theta \delta L_0}(x_2))$. Thus, since we already know that 
$s \varphi_{\theta \delta L_0}(x)$ is the image of $\varphi(x)$ under $\rho_\theta$, 
we get from formula~\eqref{wexpb} an algebraic element whose image under $\rho_{\theta}$ is
$W(\times_i \varphi_{\theta \delta L_0}(x_i))$. Once we have found those elements, 
we then get via eq.~\eqref{wexpa} algebraic elements 
$F(s; \times_i f_i \Phi_i)$ in $\cX(V, \g)$ 
whose image under $\rho_\theta$ is $s^N \, T_{\theta \delta L_0}(\prod \Phi_i(f_i))$. 

It can be shown explicitly  
that the quantities $F(s; \times_i f_i \Phi_i)$ are ($s$-dependent) local and covariant fields 
in the sense of our definition of local and covariant fields
in the free theory (see eq.~\eqref{lcf}), and that they have a smooth/analytic 
dependence on the metric under smooth/analytic variations of the metric. 
It is straightforward to show that the quantities $F(s; \times_i f_i \Phi_i)$ satisfy the causal factorization property
\begin{equation}
F(s; \times_{i=1}^n f_i \Phi_i) = F(s; \times_{i \in I} f_i \Phi_i)  \star F(s; \times_{j \in J} f_j \Phi_j) 
\end{equation}  
whenever $J^-(\supp f_i) \cap \supp f_j = \emptyset$ for all $(i, j) \in I \times J$, where
$I \cup J = \{1, \dots, n\}$ is a partition into disjoint sets.
It can be shown from eq.~\eqref{com} that the fields  $F(s; \times_i f_i \Phi_i)$ also satisfy 
the commutator property with a free field. Thus, these fields give a prescription $\tilde T(\prod \tilde \Phi_i(f_i))$
for defining time-ordered-products to which our uniqueness theorem described in section~2 can be applied\footnote{
Note however that the time-ordered-products $\tilde T(\prod \tilde \Phi_i)$ are by construction only defined as formal 
power series in $\mathcal X(V, \g)$ rather than $\cW(V, \g)$, 
since they may depend on $s$ which is itself a formal power series in $\kappa$. 
It is however not difficult to see that our uniqueness theorem can nevertheless still be applied.}. 

By this uniqueness result, the 
relation between the prescription $\tilde T$ and the original
prescription $T$ for time-ordered-products in the free theory is 
given by eq.~\eqref{SM}. This is equivalent to 
\begin{equation}
\label{ioda}
\rho_{\theta} \left[ S(\sum f_i \Phi_i) \right] = \bS_{\theta \delta L_0}
(s^{M_i} \sum f_i \Phi_i + \delta(s; \sum f_i \Phi_i)), 
\end{equation}
where the $\delta$ was introduced in eq.~\eqref{deldef}, and where $M_i$ is the number of 
factors of $\varphi$ in the field $\Phi_i$. 
(Note that $\delta$ now has an additional
$s$-dependence, due to the fact that the prescription $\tilde T$ is $s$-dependent.) 
Equation~\eqref{ioda} is the key identity for this proof. 
In order to exploit it, 
we introduce a cutoff function $\theta'$ which equals 1 
on $V$ and which is such that the support of $\theta'$ is contained in the region where $\theta$ 
equals 1. If we now apply $\rho_\theta$ to the element $\bS_{\theta' L_1}(\sum f_i \Phi_i)$, use 
eq.~\eqref{ioda} and proceed in a similar way as in the proof of eq.~\eqref{VAV} 
in section~3.2 to bring the 
resulting expression into a convenient form, then we obtain the identity 
\begin{equation}
\label{ioda1}
{\rm Ad}(V(\theta, \theta')) \circ \rho_{\theta}[\bS_{\theta' L_1}(\sum f_i \Phi_i)] = 
\bS_{\theta' K_1(s)} (\sum N_i(s) f_i \Phi_i + \bdel_{L_1}(s; \sum f_i \Phi_i)) 
\end{equation} 
for all test densities $f_i$ with support in $V$. 
Here, $V(\theta,\theta')$ is a unitary that is defined in a similar way as the unitary $W(\theta)$ in 
the proof of eq.~\eqref{VAV} in section~3.2, $N_i(s)$ are formal power series in $s$, 
$\bdel_{L_1}$ is defined as in eq.~\eqref{bdel}, and $\cK_1(s)$ is the interaction Lagrangian given by
\begin{equation}
\cK_1(s) = (s^2 - 1) \cL_0 + s^4 \cL_1 + \delta(s; \theta \cL_1) |_{\theta = 1}.
\end{equation} 
Finally, the desired *-isomorphism $\rho_s$ is then obtained from eq.~\eqref{ioda1} by removing the cutoff 
represented by $\theta$ and $\theta'$ in the same way as in our construction of the interacting field
given in section~3.1. Equation~\eqref{1pe''} corresponds to the special case $\Phi = \varphi$ of 
eq.~\eqref{ioda1}.

We finally remark that, as indicated above,  if the prescription $\tilde T$ given by eq.~\eqref{Ttilde} were actually equal 
to the original prescription $T$ for defining the time-ordered-products, then the term $\delta(s; \sum f_i\Phi_i)$
appearing in eq.~\eqref{ioda} would be zero. This would imply that the factors $N_i(s)$ in 
eq.~\eqref{ioda1} is equal to $s^{M_i}$ (where $M_i$ is the number of factors of $\varphi$ in the 
field $\Phi_i$), the term $\bdel_{L_1}(s; \sum f_i \Phi_i)$ in eq.~\eqref{ioda1} would vanish, and 
the Lagrangian $\cK_1(s)$ would be equal to $\cL_1(s)$ given by eq.~\eqref{l1s}
as in the classical theory. Thus, eq.~\eqref{1pe''} in the statement of the theorem would be simplified to
$\rho_s[\varphi_{L_1}(f)] = s \varphi_{L_1(s)}(f)$, in complete analogy with the classical theory.

\section{How to calculate the renormalization group in terms of Feynman diagrams}

In the previous sections we have set up a general framework for describing 
how a given perturbative interacting field theory in curved spacetime 
changes under a change of lengthscale, or, more properly, under a rescaling of the 
metric. This has led us to a completely satisfactory 
notion of the renormalization group flow in curved spacetime, without 
thereby having to introduce arbitrary vacuum states, bare couplings, 
cutoffs or arbitrary mass scales into the theory.  

However, our construction is rather abstract and it may not be obvious how 
one would calculate this flow in practice (to a given order in 
perturbation theory). We will now outline how this 
can be done, and we will thereby establish the connection between the 
framework explained above and the formalism of Feynman diagrams, which 
is commonly used to define the renormalization group flow in Minkowski
spacetime\footnote{We have already noted at the end of section 4.1 that the functions $G_1, G_3, G_4$
appearing in the renormalization group flow (see eq.~\eqref{RGflow}) can be determined in 
Minkowski spacetime, and they can be calculated by standard methods. However, the 
function $G_2$ must be calculated in curved spacetime.}.

To begin, we define~\cite{hw1, hw2}, for sufficiently nearby points, ``locally normal ordered'' fields
$\lno \prod \varphi^{k_i}(x_i) \rno_H$ by 
\begin{equation}
\label{lwprod}
\lno \prod_{i = 1}^n \varphi^{k_i}(x_i) \rno_H \,\, \equiv
\frac{\delta^{|k|}}{i^{|k|} \delta f(x_1)^{k_1} 
\dots \delta f(x_n)^{k_n}} \exp \left[ i\varphi(f) + \frac{1}{2}H(f, f) \right], 
\end{equation} 
where $|k| = \sum k_i$ and where 
\begin{equation}
H(x_1, x_2) = U(x_1, x_2) P(\sigma^{-1}) + V(x_1, x_2) \ln |\sigma|
\end{equation}
is the ``local Hadamard parametrix''. Since $\lno \varphi^k(x) \rno_H$ itself  
is a prescription for defining Wick powers to which our uniqueness theorem applies~\cite{hw1}, it is 
possible to expand the Wick powers $\varphi^k(x)$ in a ``local Wick expansion'' 
in terms of these locally normal ordered fields~\cite{hw1}, 
\begin{equation}
\label{98}
\varphi^k(x) = \sum_{j \le k} {k \choose j} t_{k-j}(x) \lno \varphi^j(x) \rno_H,  
\end{equation}
where $t_k$ are finite sums of terms of the form local curvature terms times parameters in the free theory, 
of the appropriate engineering dimension. Of course, 
if the prescription for defining
Wick powers is chosen to be that of ``local normal ordering'' with
respect to $H$, then the expansion of eq.~\eqref{98} is trivial, i.e., we
have $t_0 = 1$ and $t_j = 0$ for all $j >0$.
A similar expansion is possible also for the time-ordered-products~\cite{hw2},  
\begin{equation}
\label{laber}
T(\prod_{i=1}^n \varphi^{k_i}(x_i)) = \sum_{j \le k} {k \choose j} 
t_{k_1-j_1 \dots k_n-j_n}(x_1, \dots, x_n) \lno 
\prod_{i = 1}^n \varphi^{j_i}(x_i) \rno_H,  
\end{equation} 
where the $t_{j_1 \dots j_n}$ are certain distributions that are defined locally and covariantly 
in terms of the metric\footnote{However, 
it should be noted that
$t_{j_1 \dots j_n}$ is not actually a local, covariant (c-number) field in the
sense of \cite{bfv}, since one cannot give a local, covariant
prescription for how to choose the convex normal neighborhood that
enters the definition of $H$.}, and where in eq.~(\ref{fexp}) we use
the multi-index notation $j = (j_1, \dots, j_n)$, $j! = \prod_i j_i!$ etc. 

The local Hadamard parametrices $H$ appearing in eqs.~\eqref{98} and~\eqref{laber} 
could be chosen so that in Minkowski spacetime it coincides
with the symmetrized two-point function of the unique, Poincare invariant
vacuum state. In that case, when restricted to Minkowski spacetime, the
``local normal ordering'' prescription for defining Wick powers would
coincide with the (globally defined) normal ordering with respect to the
Poincare invariant vacuum state. Thus, in Minkowski spacetime, the
expansion (\ref{fexp}) could be viewed as expressing time-ordered-products
in terms of normal ordered products with repect to the usual vacuum state.
In curved spacetime, it also would be possible to choose a globally
defined ``vacuum state'' (i.e., a quasi-free Hadamard state), $\omega$,
and perform Wick expansions in terms of Wick products that are normal
ordered with respect to $\omega$. This would have the advantage that the
resulting coeficients $t$ would be globally defined rather than
being defined only on a neighborhood of the total diagonal. However, it
would have the major disadvantages that (i) the expansion (98) would
always be nontrivial (since a local, covariant field cannot coincide with
a normal ordered field on all spacetimes \cite{hw1}) and (ii) the
$t$ would no longer be locally and covariantly constructed out of
the metric, so one could not evaluate the $t$ by local computations.

The distributions $t$ can further be decomposed into contributions
from individual Feynman diagrams as follows. Let $\cF^{(k)}$ be the set of all 
Feynman diagrams consisting with $n$ vertices located at the points 
$x_1, \dots, x_n$ that are connected by a single kind of line, with 
the properties that the lines
may emerge and end on two different vertices or 
they may emerge and end on the same vertex, and the $i$th 
vertex has precisely $k_i$ edges emerging/ending on it. If $\Gamma$ is such a Feynman 
graph, then we denote by $E(\Gamma)$ the set of edges and by $V(\Gamma)$ the set 
of vertices. If $e$ is an edge, then we write $s(e)$ for the source of $e$ and 
$t(e)$ for its target. If $v$ is a vertex, then we write $n(v)$ for twice the number of 
edges that have $v$ both as their starting and endpoint. For points $x_1, \dots, x_n$ such 
that $x_i \neq x_j$ for all $i, j$, we then have
\begin{eqnarray*}
\label{fexp}
t_{k_1 \dots k_n}(x_1, \dots, x_n) &=& \sum_{\Gamma \in \cF^{(k)}} c^\Gamma \prod_{e \in E(\Gamma)}
H_F(x_{s(e)}, x_{t(e)}) \prod_{v \in V(\Gamma)} t_{n(v)}(x_v) \\
&\equiv& \sum_{\Gamma \in \cF^{(k)}} t^\Gamma(x_1, \dots, x_n), 
\end{eqnarray*} 
where $c^\Gamma$ are combinatorical factors and 
$H_F$ is the ``local Feynman parametrix'' given by 
\begin{equation}
H_F(x_1, x_2) = U(x_1, x_2) (\sigma + i0)^{-1} + V(x_1, x_2) 
\ln (\sigma + i0). 
\end{equation}
Equation~\eqref{fexp} can be viewed as giving the ``Feynman rules'' in curved spacetime. 
Mainly for simplicity, we have only considered explicitly only time-ordered-products of 
Wick powers without derivatives. Our discussion can be generalized to 
give similar Feynman rules also for time-ordered-products containing derivatives. 

The Feynman rules in curved spacetime are thus very similar to those in
Minkowski spacetime, with the local Feynman parametrix (100) replacing the
usual Feynman propagator. However, there is one key difference in that if
the prescription used for defining Wick powers does not coincide with
``local normal ordering'', then the Wick expansion (98) will be
nontrivial, and there will be correspondingly nontrivial Feynman diagrams
containing lines that begin and end at the same vertex.

The distributions $t^\Gamma$ in eq.~\eqref{fexp} are locally and covariantly constructed from the metric 
and the coupling parameters in the free theory. They 
describe the contribution of an individual Feynman graph to a time-ordered-product. Formula~\eqref{fexp}
only determines them as distributions on the product manifold $M^{\times n}$ minus the union of all of its 
partial diagonals. A prescription for the extension of all
time-ordered-products to all of $M^{\times n}$ is usually called
``renormalization''. The existence of a renormalization prescription
satisfying a list of necessary properties was proven in \cite{hw2} without
going through the intermediate step of expanding the $t_{k_1 \dots k_n}$ in terms of
Feynman diagrams.

Given the distributions $t^\Gamma$ corresponding to a given prescription $T$ for defining time ordered 
products, we can now obtain the corresponding rescaled prescription $^\lambda T$ (see eq.~\eqref{philambda})
as follows: If $p = (m^2, \xi)$ and $p(\lambda) = (\lambda^{-2} m^2, \xi)$, 
we first set
\begin{equation}
t^\Gamma_\lambda [M, \g, p] \equiv \lambda^{2|E(\Gamma)|} \cdot t^\Gamma[M, \lambda^2 \g, p(\lambda)]
\end{equation} 
as well as 
\begin{equation}
H_\lambda [M, \g, p] \equiv \lambda^{2} \cdot H[M, \lambda^2 \g, p(\lambda)].
\end{equation} 
The rescaled prescription $^\lambda T$ is then given by 
\begin{equation}
\label{fexp1}
^\lambda T(\prod_{i=1}^n {^\lambda \varphi^{k_i}}(x_i)) = 
\sum_{j \le k} \sum_{\Gamma \in \cF^{(k-j)}}
t^{\Gamma}_\lambda(x_1, \dots, x_n) \lno 
\prod_{i = 1}^n \varphi^{j_i}(x_i) \rno_{H_\lambda}.  
\end{equation} 
Given the rescaled prescription $^\lambda T$, we can now compute the 
maps $O_n(\lambda; \times_i f_i \Phi_i)$ (see eq.~\eqref{freesc}), 
which relate the rescaled prescription to the original prescription $T$.
The renormalization group flow $\cL_1 (\lambda)$ is then given in terms of these quantities by 
given by 
\begin{equation}
\label{Rn}
\delta \cL_1(\lambda) = \sum_{n=1}^\infty \frac{i^{n-1}}{n!} 
O_n(\lambda; \times^n \theta \cL_1) \Bigg|_{\theta=1}.  
\end{equation}
Each term in the sum~\eqref{Rn} is of the 
form~\eqref{countert} for some real coupling constants 
$\delta m^{2(n)}$, $\delta z^{(n)}$, $\delta \xi^{(n)}$, 
and $\delta\kappa^{(n)}$, each of which is a polynomial in $\ln \lambda$. 
These quantities are the renormalization group flow at $n$-th order in 
perturbation theory.

\medskip

This completes our brief discussion on how to calculate the renormalization group flow 
in terms of Feynman diagrams. We note, however, that 
the calculation of the $\beta$-function as defined by~\eqref{bfunction}
is more complicated since it also requires the calculation of $\cK_1(s)$ 
(see appendix A).

\end{document}